\documentclass[11pt,a4paper]{article}
\pdfoutput=1

\usepackage[colorlinks=true, linkcolor=black!50!blue, urlcolor=blue, citecolor=blue, anchorcolor=blue]{hyperref}
\usepackage[font=small,labelfont=bf,labelsep=period]{caption}
\usepackage[a4paper,top=3cm,bottom=2.5cm,left=2.5cm,right=2.5cm,bindingoffset=0mm]{geometry}

\usepackage{graphicx}
\usepackage{float}
\usepackage{afterpage}
\usepackage{epsfig}
\usepackage{cite}
\usepackage{amssymb}
\usepackage{amsmath}
\usepackage{dsfont}
\usepackage{multirow}
\usepackage{url}
\usepackage{xcolor}
\usepackage{afterpage}
\usepackage{hyperref}
\usepackage{booktabs}
\usepackage{url}
\usepackage{hyperref}
\usepackage{tabularx}

\newcolumntype{C}{>{\centering\arraybackslash}X}

\begin{document}
\begin{figure}[h]
  \includegraphics[width=0.32\textwidth]{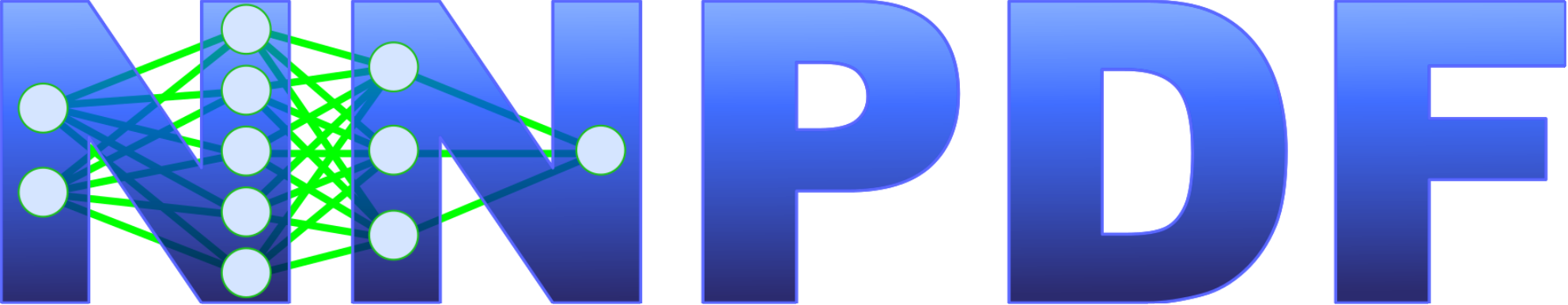}
\end{figure}
\vspace{-2.5cm}
\begin{flushright}
Edinburgh 2018/4\\
Nikhef 2018-062\\
\end{flushright}
\vspace{1cm}

\begin{center}
  {\Large \bf Nuclear Uncertainties in the Determination of Proton PDFs\\}
  \vspace{1.5cm}
  {\small
  {\bf  The NNPDF Collaboration:} \\[0.2cm]
  Richard~D.~Ball$^1$, 
  Emanuele~R.~Nocera$^{1,2}$
  and Rosalyn~L.~Pearson$^1$}

\vspace{0.5cm}
{$^1$\it \small The Higgs Centre for Theoretical Physics,\\ 
 University of Edinburgh, JCMB, KB, Mayfield Rd, Edinburgh EH9 3FD, Scotland\\
 $^2$ Nikhef Theory Group, 
 Science Park 105, 1098 XG Amsterdam, The Netherlands\\
}

\vspace{1cm}

{\bf \large Abstract}

\end{center}
We show how theoretical uncertainties due to nuclear effects may be 
incorporated into global fits of proton parton distribution functions (PDFs) 
that include deep-inelastic scattering and Drell-Yan data on 
nuclear targets. We specifically consider the CHORUS, NuTeV and E605 data 
included in the NNPDF3.1 fit, which used Pb, Fe and Cu targets, respectively. 
We show that the additional uncertainty in the proton PDFs due to nuclear 
effects is small, as expected, and in particular that the effect on the 
$\bar{d}/\bar{u}$ ratio, the total strangeness $s+\bar{s}$, and the
strange valence distribution $s-\bar{s}$ is negligible.

\section{Introduction}
\label{sec:introduction}

Modern sets of parton distribution functions  
(PDFs)~\cite{Butterworth:2015oua} are 
currently determined for the proton from a global quantum chromodynamics (QCD)
analysis of hard-scattering measurements~\cite{Gao:2017yyd}.
A variety of hadronic observables are used, including some from processes that 
do not (exclusively) involve protons in the initial state, such as 
deep-inelastic scattering (DIS) and Drell-Yan (DY) in
experiments with deuterium or heavy nuclear fixed targets.
These experiments complement the proton-only ones, providing important  
sensitivity to light PDF flavour separation~\cite{Ball:2017nwa}, and 
are therefore included in most contemporary global QCD analyses.

The inclusion of nuclear data necessitates accounting for differences between
the PDFs for free nucleons and those for partons contained within nuclei.
In the past a variety of 
different approaches have been adopted: nuclear corrections can be 
ignored on the basis that they are small~\cite{Ball:2017nwa}, 
included according to various
nuclear models~\cite{Harland-Lang:2014zoa,Dulat:2015mca,Alekhin:2017kpj}, or 
determined in a fit to the data~\cite{Accardi:2016qay}.
Whatever approach is adopted, nuclear effects will necessarily increase the  
uncertainty in the proton PDFs, since 
nuclear corrections are not known very precisely. 
So far the size of this uncertainty has also been regarded as 
small~\cite{Ball:2009mk,Ball:2013gsa}.
However, in recent years the inclusion of increasingly precise LHC 
measurements in global PDF fits has reduced PDF uncertainties to the level 
of a few percent~\cite{Ball:2017nwa}. 
Furthermore, nuclear effects have been claimed to alter the shape of 
the PDFs, especially at large values of the momentum fraction 
$x$~\cite{Owens:2012bv}, albeit without an estimate of the corresponding 
theoretical uncertainty.
Given this, it is becoming increasingly desirable to provide PDF sets 
that include such an uncertainty.

In this paper we will show how this may be achieved by performing global 
fits which include nuclear uncertainties, in the framework of the 
NNPDF3.1 global analysis~\cite{Ball:2017nwa}. We focus on the 
DIS and DY datasets with heavy nuclear targets (Pb, Fe and Cu).
We estimate the theoretical uncertainty due to neglecting the corresponding 
nuclear corrections, we include it in a fit along with the experimental
uncertainty, 
and we assess its impact on the resulting PDFs. A similar exercise for DIS 
and DY datasets with deuterium targets will be carried out in a separate 
analysis. 

Our study is accomplished within the formalism of Ref.~\cite{Ball:2018odr},
that was developed to include a broad class of theoretical uncertainties 
in a PDF fit.
The method consists of adding to the experimental covariance matrix
a theoretical covariance matrix, estimated in the space of the 
data according to the theoretical 
uncertainties associated with the theoretical predictions. 
In practice this means that the theoretical 
uncertainties are treated in much the same way as experimental systematics.
Here we will estimate the theoretical uncertainties associated with 
nuclear effects. This will be done empirically, by directly comparing 
theoretical predictions computed using nuclear PDFs (nPDFs) to 
those computed using proton PDFs. This 
comparison will allow us to construct the covariance matrix  
associated with the nuclear effects, and thus incorporate these effects 
in a global proton PDF fit. The fitting methodology will otherwise be the same 
as in the NNPDF3.1 analysis, allowing for a direct comparison of the results.

The paper is organised as follows.
In Sect.~\ref{sec:covariance}, we summarise the changes in methodology 
required to include theoretical uncertainties in a global fit. We then  
describe the nuclear dataset included in this analysis, emphasising 
to which PDF flavours it is most sensitive.
We next provide two alternative prescriptions to estimate the theoretical 
covariance matrix, and discuss their implementation.
In Sect.~\ref{sec:results}, we present the impact of nuclear data and 
theoretical corrections in a global fit of PDFs.
We compare variants of the NNPDF3.1 determination obtained by removing the 
nuclear datasets completely, or by retaining them but accounting for 
nuclear uncertainties, and nuclear corrections.
We study the fit quality and the stability of the PDFs.
Given that the nuclear dataset is mostly sensitive to sea quark PDFs,
in Sect.~\ref{sec:pheno} we assess how the $\bar{d}$-$\bar{u}$ asymmetry, 
and the strangeness content of the proton, including the asymmetry between $s$
and $\bar{s}$ PDFs, are affected by nuclear uncertainties.
We provide our conclusions and an outlook in Sect.~\ref{sec:conclusions}.

\section{Theoretical Uncertainties due to Nuclear Corrections}
\label{sec:covariance}

In this section we describe how the NNPDF methodology can be adapted 
to incorporate theoretical uncertainties through a theoretical 
covariance matrix, as proposed in Ref.~\cite{Ball:2018odr}.
We then focus on the DIS and DY datasets in the NNPDF3.1 global dataset
(described in detail in 
Sect.~2 of Ref.~\cite{Ball:2017nwa}) that involve nuclear targets 
other than deuterium. We first summarise the type of measured observables, 
their kinematic coverage, and their sensitivity to the underlying PDFs. We 
then show how the size of the nuclear effects may be estimated empirically by 
comparing the different theoretical predictions made with proton and nuclear
PDFs.
We finally offer two alternative prescriptions to estimate the theoretical
covariance matrix for the nuclear uncertainties: the first of which simply
provides a conservative estimate of the overall uncertainty; 
the second of which also applies a correction for nuclear effects,
aiming to reduce the overall uncertainty.

\subsection{Theoretical Uncertainties in PDF Fits}
\label{subsec:thunc}

The NNPDF fitting methodology \cite{Ball:2008by} works 
in two stages. We start from a set of experimental data points 
$D_i$, $i=1,\ldots,N_{\rm dat}$, with an associated experimental covariance 
matrix $C_{ij}$ (which includes all experimental statistical and systematic 
uncertainties). We then generate data replicas 
$D_i^{(k)}$, $k=1,\ldots,N_{\rm rep}$,
which are Gaussianly distributed in such a way that their ensemble averages 
reproduce the data and their uncertainties: 
\begin{equation}
\langle D_i^{(k)}\rangle = D_i,\qquad 
\langle(D_i^{(k)}-D_i)(D_j^{(k)}-D_j)\rangle=C_{ij}\, ,
\label{eq:datagen}
\end{equation}
where the ensemble average $\langle\cdot\rangle$ is taken over a 
sufficiently large ensemble of data replicas (in principle strict equality 
only holds in the limit $N_{\rm rep}\to\infty$). 

We then compare theoretical predictions 
$T_i[f]$, which depend on the PDFs $f$ (or more precisely the neural network 
parameters which parametrise these PDFs), to the data replicas, by optimising 
a figure of merit: 
\begin{equation}
  \label{eq:chi2}
  \chi^2[f,D] = \frac{1}{N_{\rm dat}}
  \sum_{i,j} (T_{i}[f] - D_{i}) \, ( C_0^{-1})_{ij} \,
  (T_{j}[f] - D_{j} )\, .
\end{equation}
Here, $C_{0}$ is the $t_0$-covariance matrix used in the fit. If all the 
experimental uncertainties were additive, $(C_0)_{ij}$ would simply be the 
experimental covariance matrix $C_{ij}$, but in the presence of 
multiplicative uncertainties this would bias the fit~\cite{DAgostini:2003syq}. 
To eliminate this bias we use instead $(C_0)_{ij}$, constructed from $C_{ij}$
using the $t_0$ method~\cite{Ball:2009qv}. The method requires the PDF
dependence implicit in $(C_0)_{ij}$ to be iterated to consistency; in 
practice this iteration converges very rapidly, as the dependence is very weak. 

In this way we obtain a PDF replica $f^{(k)}$ for each data replica $D^{(k)}$. 
Since the distribution of the PDF replicas is a representation of 
the distribution of the data replicas, the ensemble of PDF replicas 
$\{f^{(k)}\}$ gives us a representation of the PDFs and their 
correlated uncertainties.

We can incorporate theory uncertainties into the 
NNPDF methodology by supplementing the experimental covariance matrix 
$C_{ij}$ with a theoretical covariance matrix $S_{ij}$, estimated using
the theoretical uncertainties associated with the theoretical predictions 
$T_i[f]$. The experimental and theoretical uncertainties are by their 
nature independent. In Ref.\cite{Ball:2018odr} it was shown that if we assume 
that both experimental and theoretical uncertainties are independent and 
Gaussian, the two covariance matrices can simply be added; the combined 
covariance matrix $C_{ij}+S_{ij}$ then gives the total uncertainty in the 
extraction of PDFs from the experimental data.

In practice this means that, when we generate the data replicas, in place of 
Eq.~(\ref{eq:datagen}) we need 
\begin{equation}
\langle D_i^{(k)}\rangle = D_i,\qquad 
\langle(D_i^{(k)}-D_i)(D_j^{(k)}-D_j)\rangle=C_{ij}+S_{ij},
\label{eq:datagenth}
\end{equation}
to ensure that the theoretical uncertainty is propagated through to the 
PDFs along with the experimental uncertainties. Likewise when we fit, 
in place of Eq.(\ref{eq:chi2}) we use as the figure of merit 
\begin{equation}
  \label{eq:chi2th}
  \chi^2[f,D] = \frac{1}{N_{\rm dat}}
  \sum_{i,j} (T_{i}[f] - D_{i}) \, ( C_0+S)^{-1}_{ij} \,
  (T_{j}[f] - D_{j} )\, .
\end{equation}
This ensures that the fitting accounts for the relative weight of the data 
points according to both the experimental and the theoretical uncertainties. 

It remains to estimate the theoretical covariance matrix $S_{ij}$. In general 
this will be constructed in the same way as an experimental systematic: a 
range of theoretical predictions $T_i^{(n)}$ can be characterised by nuisance 
parameters, $\Delta_i^{(n)}=T_i^{(n)}-T_i$, $n=1,\ldots,N_{\rm nuis}$. Assuming 
that we can model the theoretical uncertainties by a Gaussian characterised by 
these nuisance parameters, we can write    
\begin{equation}
S_{ij}= \mathcal{N}\sum_{n=1}^{N_{\rm nuis}}\Delta^{(n)}_i\Delta^{(n)}_j\,,
\label{eq:nuisance}
\end{equation}
where ${\cal N}$ is a normalisation which depends on whether the nuisance 
parameters are independent uncertainties, or different estimates of 
the same uncertainty. 

Note that since the predictions 
$T_i[f]$ depend on the PDFs, this means the nuisance parameters
$\Delta_i^{(n)}$ and 
the theoretical covariance matrix $S_{ij}$ will  
also depend implicitly on the PDF, albeit weakly. This can be dealt with 
in precisely the same way as in the $t_0$ method~\cite{Ball:2009qv}: $S_{ij}$ 
is computed with an initial (central) PDF, which is then iterated to 
consistency. In practice the iterations can be performed simultaneously. 

In this paper we will show how this procedure works by estimating 
the theoretical uncertainties due specifically to nuclear effects related 
to the use of data from scattering off heavy nuclear targets. 

\subsection{The Nuclear Dataset}
\label{subsec:dataset}

The NNPDF3.1 dataset involving heavy nuclei consists of 
inclusive charged-current DIS cross sections from CHORUS~\cite{Onengut:2005kv},
DIS dimuon cross sections from NuTeV~\cite{Goncharov:2001qe,Mason:2006qa},
and DY dimuon cross sections from E605~\cite{Moreno:1990sf}. 
In the case of CHORUS and NuTeV, neutrino and antineutrino beams are scattered
off a lead ($^{208}_{\ 82}$Pb) and an iron ($^{56}_{26}$Fe) target, respectively;
while in the case of E605 a proton beam is scattered off a copper 
($^{64}_{32}$Cu) target.
Henceforth we refer to the combined measurements from CHORUS, NuTeV and E605
as the nuclear dataset. In NNPDF3.1 there are additional DIS and DY datasets
involving scattering from deuterium: nuclear corrections to these datasets
will be considered in a future analysis. In NNPDF we do not use the 
CDHSW neutrino-DIS data~\cite{Berge:1989hr}, taken with an iron target.

An overview of the nuclear dataset is presented in Table~\ref{tab:dataset}, 
where we indicate, for each dataset: the observable,
the corresponding reference, the number of data points before and after
kinematic cuts, and the kinematic range covered in the relevant
variables after cuts.
Kinematic cuts match the next-to-next-to-leading order (NNLO)  
NNPDF3.1 baseline fit: for DIS we require $Q^2\geq 3.5$ GeV$^2$ 
and $W^2\geq 12.5$ GeV$^2$, where $Q^2$ and $W^2$ are the energy transfer and 
the invariant mass of the final state in the DIS process, respectively; 
for DY, we require $\tau\leq 0.080$ and $|y_{\ell\ell}/y_{\rm max}|\leq 0.663$, 
where $\tau=M_{\ell\ell}^2/s$ and $y_{\rm max}=-\frac{1}{2}\ln\tau$, with 
$y_{\ell\ell}$ and $M_{\ell\ell}$ the rapidity and the invariant mass of 
the dimuon pair, respectively, and $\sqrt{s}$ the centre-of-mass energy 
of the DY process. 

\begin{figure}[p]
\centering
\includegraphics[scale=0.95]{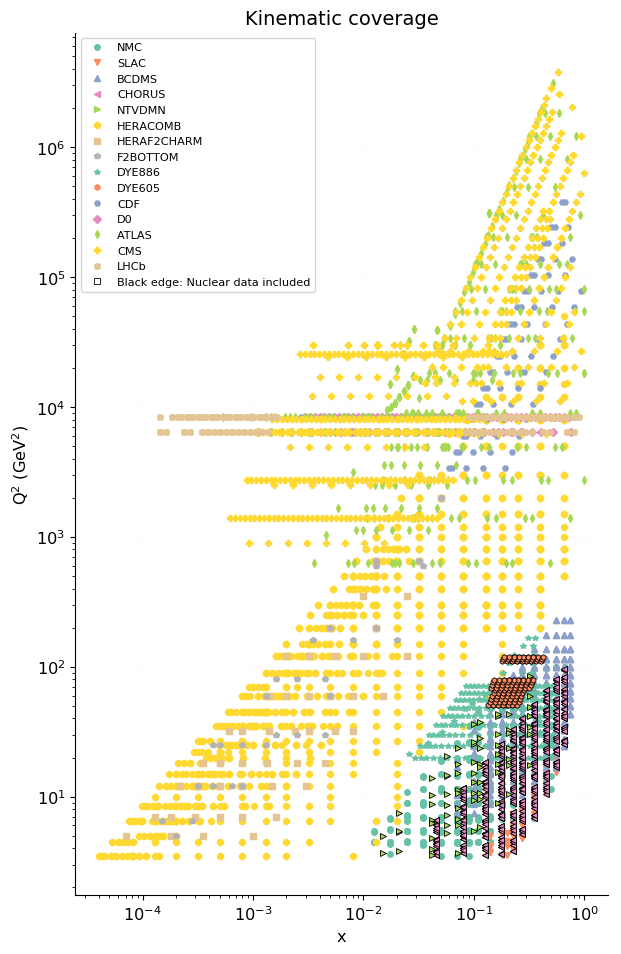}\\
\caption{The kinematic coverage of the current analysis, equivalent to that 
  of the NNPDF3.1 analysis, in the $(x,Q^2)$ plane. Points belonging to the
  nuclear dataset are circled in black.}
\label{fig:dataset}
\end{figure}

The kinematic coverage of the three experiments in the $(x,Q^2)$ plane 
is compared to the whole of the NNPDF3.1 dataset in Fig.~\ref{fig:dataset}, 
where points corresponding to the nuclear dataset are circled in black.
For hadronic data, the momentum fraction $x$ has been reconstructed from 
leading-order (LO) kinematics (using central rapidity for those observables 
integrated over rapidity), and the value of $Q^2$ has been set equal to the 
characteristic scale of the process.
The number of data points after cuts is $4285$, 
out of which $993$ belong to the nuclear
dataset (corresponding to about $23\%$ of the entire dataset). 
Most of the nuclear data (around $84\%$) are from CHORUS.

\begin{table}[!t]
\centering
\scriptsize
\begin{tabularx}{\textwidth}{lCcrrCC}
\toprule
  Experiment 
& Obs. 
& Ref.
& \multicolumn{2}{c}{$N_{\rm dat}$} 
& ${\rm Kin}_1$ 
& ${\rm Kin}_2$ [GeV] \\ 
\midrule
  \multirow{2}{*}{CHORUS} 
& $\sigma^{\nu}$ 
& \cite{Onengut:2005kv}
& 607 & (416)
& $0.045\leq x\leq 0.65$
& $1.9\leq Q\leq 9.8$ \\
& $\sigma^{\bar\nu}$ 
& \cite{Onengut:2005kv}
& 607 
& (416)
& $0.045\leq x\leq 0.65$
& $1.9\leq Q\leq 9.8$ \\
\midrule
  \multirow{2}{*}{NuTeV}
& $\sigma^{\nu,c}$ 
& \cite{Goncharov:2001qe,Mason:2006qa}
& 45 
& (39)
& $0.02\leq x\leq 0.33$
& $2.0\leq Q \leq 10.8$ \\
& $\sigma^{\bar\nu,c}$ 
& \cite{Goncharov:2001qe,Mason:2006qa}
& 45 
& (37)
& $0.02\leq x\leq 0.21$
& $1.9\leq Q \leq \ \, 8.3$ \\
\midrule
  E605
& $\sigma_{\rm DY}^p$ 
& \cite{Moreno:1990sf}
& 119 
& (85)
& $-0.2\leq y_{\ell\ell}\leq 2.9$
& $7.1\leq M_{\ell\ell}\leq 10.9$ \\
\bottomrule
\end{tabularx}
\caption{The datasets involving nuclear targets other than deuterium in 
  NNPDF3.1. The kinematic range covered in each variable is given after cuts 
  are applied (see the text for details).}
\label{tab:dataset}
\end{table}

The observables measured by CHORUS, NuTeV and E605 allow one to control the 
valence-sea (or quark-antiquark) separation at medium-to-high values of the 
momentum fraction $x$, and at a rather low energy $Q^2$.
Following their factorised form, charged current DIS cross sections measured 
by CHORUS are expected to provide some information on the valence distributions
$u_V=u-\bar{u}$ and $d_V=d-\bar{d}$; DIS dimuon cross sections, reconstructed 
by NuTeV from the decay of a charm quark, are sensitive to $s$ and 
$\bar{s}$ PDFs; and DY dimuon cross sections measured by E605 probe
$\bar{u}$ and $\bar{d}$ PDFs.

\begin{figure}[t!]
\centering
\includegraphics[width=\textwidth,clip=true,trim=3cm 0 3cm 0]{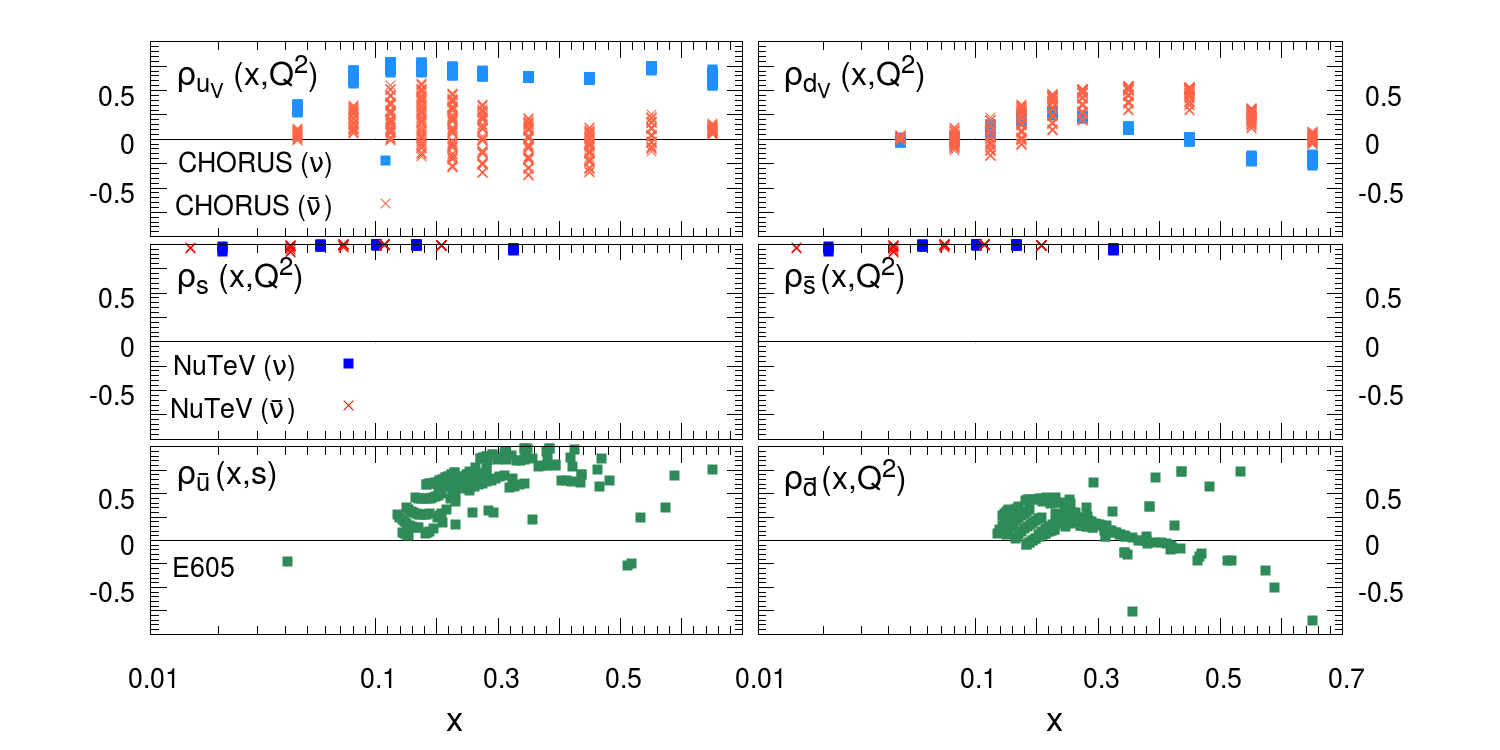}\\
\caption{The correlation coefficient $\rho$ between the observables in 
  Table~\ref{tab:dataset} and (from top to bottom) the $u_V$ and $d_V$ 
  PDFs for CHORUS, the $s$ and $\bar s$ PDFs 
  for NuTeV, and the $\bar u$ and $\bar d$ PDFs for E605.}
\label{fig:rho}
\end{figure}

The sensitivity of the measured observables to different PDF flavours can be 
quantified by the correlation coefficient $\rho$ (defined in Eq.~(1) in 
Ref.~\cite{Guffanti:2010yu}) between the PDFs in a given 
set and the theoretical predictions corresponding to the measured data points.
Large values of $|\rho|$ indicate that the sensitivity of the PDFs to the data 
is most significant.
The correlation coefficient $\rho$ is displayed in Fig~\ref{fig:rho},
from top to bottom, for the $u_V$ and $d_V$ PDFs from CHORUS,
for the $s$ and $\bar s$ PDFs from NuTeV, and for the $\bar u$ and $\bar d$ 
PDFs from E605.
Each point corresponds to a different datum in the experiments enumerated in 
Table~\ref{tab:dataset}: PDFs are taken from the NNDPF3.1 NNLO parton set,
and are evaluated at a scale equal to either the momentum transfer $Q^2$ 
(for DIS) or the center-of-mass energy $s$ (for DY) of that point.
For DY, the value of $x$ is computed from hadronic variables using LO
kinematics.
As anticipated, the correlation between the PDF flavours and the observables
displayed in Fig.~\ref{fig:rho} is sizeable, in particular: between $u_V$  
($d_V$) PDFs and the neutrino (antineutrino) charged current DIS cross 
sections from CHORUS in the range $0.1\leq x\leq 0.7$ ($0.2\leq x \leq 0.5$); 
between $s$ ($\bar s$)
PDFs and the antineutrino dimuon DIS cross sections from NuTeV along all the 
measured range, $0.02\leq x\leq 0.32$ ($0.02\leq x\leq 0.21$); 
and between $\bar{u}$ ($\bar d$) PDFs 
and the dimuon DY cross sections from E605 in the range $0.3\leq x\leq 0.7$.
Correlations between the measured observables and other PDFs, not displayed in 
Fig.~\ref{fig:rho}, are relatively small.
We therefore expect that including theoretical uncertainties
due to nuclear corrections will mainly affect the valence-sea PDF flavour
separation in the kinematic region outlined above.

\subsection{Determining Correlated Nuclear Uncertainties}
\label{subsec:covmat}

In Sec.~\ref{subsec:thunc}, we explained how, if we want to include theoretical 
uncertainties in a PDF fit, we first need to estimate the theoretical 
covariance matrix $S_{ij}$ in the space of the data, 
using Eq.~\eqref{eq:nuisance}.
In this section, we illustrate how we might achieve this for the nuclear 
uncertainties affecting the three datasets described in 
Sec.~\ref{subsec:dataset}.
First we provide two alternative definitions for the theoretical covariance
matrix associated with nuclear uncertainties, then we describe how we 
can implement them in practice, and finally we discuss the results.

\subsubsection{Definition}
\label{subsubsec:definition}

We construct the point-by-point correlated elements $S_{ij}$ of the 
theoretical covariance matrix as the unweighted average over 
$N_{\rm nuis}$ nuisance parameters $\Delta_{i,j}^{(n)}$ for each data 
point $i,j=1,\dots, N_{\rm dat}$, as in Eq.~(\ref{eq:nuisance}).
A conservative definition of the nuisance parameters for nuclear 
uncertainties, which takes into 
account all the uncertainty due to the difference between nuclear and proton 
targets, is
\begin{equation}
 \Delta_{i}^{(n)}
 = T^N_i[f_N^{(n)}]- T^N_i[f_p] \,,
 \label{eq:def1}
\end{equation}
where $T^N_i[f_N^{(n)}]$ and $T^N_i[f_p]$ denote
the theoretical prediction for the nuclear observable 
using a PDF $f_N^{(n)}$ for a heavy nucleus $N$, 
and the corresponding prediction using a proton PDF $f_p$.
The subscript $N$ identifies the appropriate isotope, {\it i.e.}
$N={^{208}_{\ 82}{\rm Pb}}$ for CHORUS, $N={^{56}_{26}{\rm Fe}}$ for NuTeV, 
and $N={^{64}_{32}{\rm Cu}}$ for E605.
The superscript $n$ identifies a particular model of nuclear corrections.

Various models of nuclear effects on PDFs exist in the literature (for a review, see, {\it e.g.}, Ref.~\cite{Arneodo:1992wf}).
They are however based on a range of different assumptions, which often limit
their validity.
In our opinion, a better ansatz for nuclear effects, over all the kinematic 
range covered by the measurements in Table~\ref{tab:dataset}, is provided by 
global fits of nPDFs, since they are primarily driven by the
data.
The $N_{\rm nuis}$ models in Eq.~\eqref{eq:nuisance} can then be identified with 
different members of a nPDF set.
The free proton PDF can instead be taken from a global set of proton PDFs: it 
should in any case be iterated to consistency at the end of the fitting 
procedure, as explained at the end of Sec.\ref{subsec:thunc}. 
The practical way in which nPDF members are constructed, 
the proton PDF is chosen, and the corresponding observables are computed
is discussed in Sect.~\ref{subsubsec:implementation} below.

A more ambitious definition of the nuisance parameters $\Delta_{i}^{(n)}$ in 
Eq.~\eqref{eq:nuisance} is to consider the theoretical uncertainty to 
be due only to the uncertainties in the nPDFs themselves. Therefore
\begin{equation}
\Delta_{i}^{(n)}
 = T^N_i[f_N^{(n)}]- T^N_i[f_N] \,,
 \label{eq:def2a}
\end{equation}
where, in comparison to Eq.~\eqref{eq:def1}, the expectation value of the 
proton observable is now replaced by the central value of the nPDFs
$f_N=\langle f_N^{(n)}\rangle$.
Because Eq.~\eqref{eq:def2a} does not contain any information on how nuclear
observables differ from the corresponding proton ones, 
Eq.~\eqref{eq:def2a} must be supplemented with a shift, applied 
to each data point $i$, that takes into account the difference between the two: 
\begin{equation}
 \delta T^N_i
 = T^N_i[f_N] - T^N_i[f_p].
\label{eq:def2b}
\end{equation}
This can be thought of as the nuclear correction to the theoretical 
predictions, which is equivalent to a correction to the data, when 
used in Eq.~(\ref{eq:chi2th}).

The two definitions are in principle different.
In Eq.~\eqref{eq:def1}, the contribution of the nuclear data to the global 
fit is {\it deweighted} by an extra uncertainty, which encompasses both the 
difference between the proton and nuclear PDFs, and the uncertainty in the 
nPDFs.
In Eqs.~\eqref{eq:def2b} the theory is {\it corrected} by a shift. Here the 
uncertainty, and thus the deweighting, is correspondingly smaller, arising 
only from the uncertainty in the nPDFs. In principle, if the uncertainty 
in the nPDFs is correctly estimated, and smaller than the shift, the second 
definition should give more precise results. However if the shift is small, 
or unreliably estimated, the first definition will be better, and should 
result in a lower $\chi^2$ for the nuclear data, albeit with slightly larger 
PDF uncertainties.

\subsubsection{Implementation}
\label{subsubsec:implementation}

The goal of our exercise is to estimate the overall level of theoretical 
uncertainty associated to nuclear effects.
Inconsistency (following from somewhat inconsistent parametrisations) should be 
part of that, therefore, instead of relying on a single nPDF determination in 
Eqs.~(\ref{eq:def1}-\ref{eq:def2b}), we find it useful to utilise a combination 
of different nPDF sets.
Such a combination can be realised in a statistically sound way by following 
the methodology developed in Ref.~\cite{Butterworth:2015oua}, which
consists in taking the unweighted average of the nPDF sets.
The simplest way of realing it is to generate equal numbers of Monte Carlo
replicas from each input nPDF set, and then merge them together in a single
Monte Carlo ensemble.
The appropriate normalisation in Eq.~\eqref{eq:nuisance} is therefore 
${\cal N} =\frac{1}{N_{\rm nuis}}$, since each replica is equally probable; each
nPDF member in Eqs.~(\ref{eq:def1}-\ref{eq:def2b}) is a replica in the Monte
Carlo ensemble; and $f_N=\langle f_N^{(n)}\rangle$ is the zero-th replica in the 
same Monte Carlo ensemble.
The combination method of Ref.~\cite{Butterworth:2015oua} has proven to be 
adequate when results are compatible or differences are understood, as is 
the case with nPDFs.

The Monte Carlo ensemble utilised to 
compute Eqs.~(\ref{eq:def1}-\ref{eq:def2b}) is determined as follows.
We consider recent nPDF sets available in the literature, namely
DSSZ12~\cite{deFlorian:2011fp}, nCTEQ15~\cite{Kovarik:2015cma}
and EPPS16~\cite{Eskola:2016oht}.
These are determined at next-to-leading order (NLO) from a global analysis 
of measurements in DIS, DY and proton-nucleus ($pN$) collisions.
A compilation of the data included in these sets is given in 
Table~\ref{tab:fullnucdata}. A detailed description may be found in  
Refs.~\cite{deFlorian:2011fp,Kovarik:2015cma,Eskola:2016oht}, and a critical 
comparison is documented, {\it e.g.}, 
in Refs.~\cite{Paukkunen:2017bbm,Paukkunen:2018kmm}.
As one can see, all three determinations include a significant
amount of experimental information, so should collectively provide a 
reasonable representation of nuclear modifications.

The nuclear datasets used in NNPDF3.1, and hence in the fits 
performed in this analysis, also enter some of the nPDFs selected above.
This is the case of NuTeV measurements (also included in DSSZ12) and 
of CHORUS measurements (also included in DSSZ12 and EPPS16), see
Table~\ref{tab:dataset} and Table~\ref{tab:fullnucdata}.
This does not lead to a double counting of these data because we only use 
the nPDFs as a model to establish an additional correlated source of 
uncertainty in the determination of the proton PDFs. This then leads to an 
increase in overall uncertainties, since the nuclear datasets are deweighted, 
while double counting would give a decrease in uncertainties. 

However the nuclear corrections, computed in this way, implicitly assume
an underlying proton PDF (this is what a fit of nPDFs does).
In this respect, the process is conceptually equivalent to 
the inclusion of nuclear corrections in a fit of proton PDFs according 
to some phenomenological model whose parameters are tuned to 
the data beforehand, as done, for instance,
in Ref.~\cite{Harland-Lang:2014zoa}. In principle, the procedure should be 
iterated to consistency \cite{Ball:2018odr}: the output of a proton PDF 
fit including 
the nuclear uncertainties can be used to update a fit of nPDFs, which can 
be used in turn to refine the estimate of the theoretical covariance 
matrix, Eq.~\eqref{eq:nuisance}, to be included in a subsequent fit of 
proton PDFs.
In practice however this iteration is unecessary, since we will find that 
the effect of the nuclear correction is already very small. This is fortunate, 
since we are in any case not yet able to consistently perform nPDF fits 
within the NNPDF framework.
%

\begin{table}[!p]
\centering
\scriptsize
\renewcommand{\arraystretch}{1.13}
\begin{tabularx}{\textwidth}{XXcccc}
\toprule
  Observable 
& Experiment 
& Ref. 
& $N_{\rm dat}$ (DSSZ12)  
& $N_{\rm dat}$ (nCTEQ15) 
& $N_{\rm dat}$ (EPPS16) \\
\midrule
  $F_2^{\rm D}/F_2^{\rm D}$ 
& NMC     & \cite{Arneodo:1996qe}                 & --- & 201 & --- \\
  $F_2^{\rm He}/F_2^{\rm D}$ 
& NMC     & \cite{Amaudruz:1995tq}                &  17 &  12 &  16 \\
& E139    & \cite{Gomez:1993ri}                   &  18 &   3 &  21 \\
& HERMES  & \cite{Airapetian:2002fx}              & --- &  17 & --- \\
  $F_2^{\rm Li}/F_2^{\rm D}$
& NMC     & \cite{Arneodo:1995cs}                 &  17 &  11 &  15 \\
  $F_2^{\rm Li}/F_2^{\rm D}$ $Q^2$ dep
& NMC     & \cite{Arneodo:1995cs}                 & --- & --- & 153 \\
  $F_2^{\rm Be}/F_2^{\rm D}$
& E139    & \cite{Gomez:1993ri}                   &  17 &   3 &  20 \\
  $F_2^{\rm C}/F_2^{\rm D}$
& E665    & \cite{Adams:1995is}                   & --- &   3 & --- \\
& E139    & \cite{Gomez:1993ri}                   &   7 &   2 &   7 \\
& EMC     & \cite{Ashman:1988bf}                  &   9 &   9 & --- \\
& NMC     & \cite{Arneodo:1996qe,Amaudruz:1995tq} &  17 &  24 &  31 \\
  $F_2^{\rm C}/F_2^{\rm D}$ $Q^2$ dep
& NMC     & \cite{Amaudruz:1995tq}                & 191 & --- & 165 \\
  $F_2^{\rm N}/F_2^{\rm D}$
& HERMES  & \cite{Airapetian:2002fx}              & --- &  19 & --- \\
& BCDMS   & \cite{Bari:1985ga}                    & --- &   9 & --- \\
  $F_2^{\rm Al}/F_2^{\rm D}$
& E139    & \cite{Gomez:1993ri}                   &  17 &   3 &  20 \\
  $F_2^{\rm Ca}/F_2^{\rm D}$
& NMC     & \cite{Amaudruz:1995tq}                &  16 &  12 &  15 \\
& E665    & \cite{Adams:1995is}                   & --- &   3 & --- \\
& E139    & \cite{Gomez:1993ri}                   &   7 &   2 &   7 \\
  $F_2^{\rm Fe}/F_2^{\rm D}$
& E049    & \cite{Bodek:1983qn}                   & --- &   2 & --- \\
& E139    & \cite{Gomez:1993ri}                   &  23 &   6 &  26 \\
& BCDMS   & \cite{Benvenuti:1987az,Bari:1985ga}   & --- &  16 & --- \\
  $F_2^{\rm Cu}/F_2^{\rm D}$
& EMC     & \cite{Ashman:1992kv,Ashman:1988bf}    &  19 &  27 &  19 \\
  $F_2^{\rm Kr}/F_2^{\rm D}$
& HERMES  &  \cite{Airapetian:2002fx}             & --- &  12 & --- \\
  $F_2^{\rm Ag}/F_2^{\rm D}$
& E139    &  \cite{Gomez:1993ri}                  &   7 &   2 &   7 \\
  $F_2^{\rm Sn}/F_2^{\rm D}$
& EMC     &   \cite{Ashman:1988bf}                &   8 &   8 & --- \\
  $F_2^{\rm Xe}/F_2^{\rm D}$
& E665    & \cite{Adams:1995is}                   & --- &   2 & --- \\
  $F_2^{\rm Au}/F_2^{\rm D}$
& E139    &  \cite{Gomez:1993ri}                  &  18 &   3 &  21 \\
  $F_2^{\rm Pb}/F_2^{\rm D}$
& E665    &  \cite{Adams:1995is}                  & --- &   3 & --- \\
  $F_2^{\rm C}/F_2^{\rm Li}$
& NMC     & \cite{Amaudruz:1995tq}                &  24 &   7 &  20 \\
  $F_2^{\rm Ca}/F_2^{\rm Li}$
& NMC     & \cite{Amaudruz:1995tq}                &  24 &   7 &  20 \\
  $F_2^{\rm Be}/F_2^{\rm C}$
& NMC     & \cite{Arneodo:1996rv}                 &  15 &  14 &  15 \\
  $F_2^{\rm Al}/F_2^{\rm C}$
& NMC     & \cite{Arneodo:1996rv}                 &  15 &  14 &  15 \\
  $F_2^{\rm Ca}/F_2^{\rm C}$
& NMC     & \cite{Amaudruz:1995tq}                &  39 &  21 &  15 \\
  $F_2^{\rm Fe}/F_2^{\rm C}$
& NMC     & \cite{Arneodo:1996rv}                 &  15 &  14 &  15 \\
  $F_2^{\rm Sn}/F_2^{\rm C}$
& NMC     & \cite{Arneodo:1996ru}                 &  15 & --- &  15 \\
  $F_2^{\rm Sn}/F_2^{\rm C}$ $Q^2$ dep
& NMC     & \cite{Arneodo:1996ru}                 & 145 & 111 & 144 \\
  $F_2^{\rm Pb}/F_2^{\rm C}$
& NMC     & \cite{Arneodo:1996rv}                 &  15 &  14 &  15 \\
  $F_2^{\rm \nu Fe}$
& NuTeV   & \cite{Tzanov:2005kr}                  &  75 & --- & --- \\
& CDHSW   & \cite{Berge:1989hr}                   & 120 & --- & --- \\
  $F_3^{\rm \nu Fe}$
& NuTeV   & \cite{Tzanov:2005kr}                  &  75 & --- & --- \\
& CDHSW   & \cite{Berge:1989hr}                   & 133 & --- & --- \\
  $F_2^{\rm \nu Pb}$
& CHORUS  & \cite{Onengut:2005kv}                 &  63 & --- & 412 \\
  $F_3^{\rm \nu Pb}$
& CHORUS  & \cite{Onengut:2005kv}                 &  63 & --- & 412 \\
\midrule
  $\sigma_{\rm DY}^{\rm pC}/\sigma_{\rm DY}^{\rm pD}$
& E772    & \cite{Alde:1990im}                    &   9 &   9 &   9 \\
  $\sigma_{\rm DY}^{\rm pCa}/\sigma_{\rm DY}^{\rm pD}$
& E772    & \cite{Alde:1990im}                    &   9 &   9 &   9 \\
  $\sigma_{\rm DY}^{\rm pFe}/\sigma_{\rm DY}^{\rm pD}$
& E772    & \cite{Alde:1990im}                    &   9 &   9 &   9 \\
  $\sigma_{\rm DY}^{\rm pW}/\sigma_{\rm DY}^{\rm pD}$
& E772    & \cite{Alde:1990im}                    &   9 &   9 &   9 \\
  $\sigma_{\rm DY}^{\rm pFe}/\sigma_{\rm DY}^{\rm pBe}$
& E886    & \cite{Vasilev:1999fa}                 &  28 &  28 &  28 \\
  $\sigma_{\rm DY}^{\rm pW}/\sigma_{\rm DY}^{\rm pBe}$
& E886    & \cite{Vasilev:1999fa}                 &  28 &  28 &  28 \\
\midrule
  $d\sigma_{\pi^0}^{\rm dAu}/d\sigma_{\pi^0}^{pp}$
& PHENIX  & \cite{Adler:2006wg}                   &  21 &  20 &  20 \\
& STAR    & \cite{Abelev:2009hx}                  & --- &  12 & --- \\
  $d\sigma_{\pi^-}^{\rm pW}/d\sigma_{\pi^-}^{pD}$
& NA10    & \cite{Bordalo:1987cs}                 & --- & --- &  10 \\
  $d\sigma_{\pi^+}^{\rm pW}/d\sigma_{\pi^-}^{pW}$
& E615    & \cite{Heinrich:1989cp}                & --- & --- &  11 \\
  $d\sigma_{\pi^-}^{\rm pPt}/d\sigma_{\pi^-}^{pH}$
& NA3     & \cite{Badier:1981ci}                  & --- & --- &   7 \\
  $W^{\rm \pm}$ pPb
& CMS     & \cite{Khachatryan:2015hha}            & --- & --- &  10 \\
  $Z$ pPb
& CMS     & \cite{Khachatryan:2015pzs}            & --- & --- &   6 \\
& ATLAS   & \cite{Aad:2015gta}                    & --- & --- &   7 \\
  dijet pPb
& CMS     & \cite{Sassot:2009sh}                  & --- & --- &   7 \\
\midrule
&         &                                       &1579 &1627 &1811 \\
\bottomrule
\end{tabularx}
\caption{The DIS, DY and $pp$ data 
   entering the nPDF sets used in this analysis.}
\label{tab:fullnucdata}
\end{table}


The three nPDF sets selected above are each delivered as Hessian sets, 
corresponding to 90\% confidence levels (CLs) for nCTEQ15 and EPPS16.
These CLs were determined by requiring a tolerance $T=\sqrt{\Delta\chi^2}$,
with $\Delta\chi^2=35$, and $\Delta\chi^2=52$, for the two nPDF sets, 
respectively.
Excursions of the individual eigenvector directions resulting in a 
$\Delta\chi^2$ up to 30 units, which correspond to an increase in $\chi^2$ 
of about $2\%$, are tolerated in DSSZ12.
We assume that this represents the $68\%$ CL of the fit~\cite{Sassot:PC}.
To generate corresponding Monte Carlo sets, we utilise the Thorne-Watt 
algorithm~\cite{Watt:2012tq} implemented in the public code of 
Ref.~\cite{Hou:2016sho}, and rescale all uncertainties to 68\% CLs.
We then generate $300$ replicas for each nPDF set.
The size of the Monte Carlo ensemble is chosen to reproduce the central value 
and the uncertainty of the original Hessian sets with an accuracy 
of few percent. 
Such an accuracy is smaller than the spread of the nPDF sets, 
and therefore adequate for our purpose.
We assume that all the three nPDF Monte Carlo sets are equally likely 
representations of the same underlying probability distribution, 
and we combine them by choosing equal numbers of replicas from 
each set.
In the case of CHORUS and NuTeV, both lead and iron nPDFs are available from 
DSSZ12, nCTEQ15, and EPPS16, therefore the total number of replicas in 
the combined set is $N_{\rm nuis}=900$; in the case of E605, 
copper nPDFs are only available from nCTEQ15 and EPPS16, 
therefore for E605 $N_{\rm nuis}=600$.
In principle, the large number of Monte Carlo replicas (and nuisance 
parameters) can be reduced by means of a suitable compression 
algorithm~\cite{Carrazza:2015hva} without a significant statistical loss.
A Monte Carlo ensemble of approximately the typical size of the starting 
Hessian sets could thus be obtained. However, we do not find it necessary 
to do this in the current analysis.

We then use this Monte Carlo ensemble of nPDFs, $\{\tilde{f}_{p/N}^{(n)}\}$, 
to determine the nuclear correction factors
\begin{equation}
R_f^{N,(n)}=\frac{\tilde{f}_{p/N}^{(n)}}{\tilde{f}_p}\,,
\label{eq:ratio}
\end{equation}
where $\tilde{f}_{p/N}^{(n)}$ is the bound-proton replica and $\tilde{f}_p$ 
is the central value of the free-proton PDF originally used in the 
corresponding nPDF analysis (namely CT14~\cite{Dulat:2015mca} for EPPS16, 
a variation of the CTEQ6.1 analysis presented in Ref.~\cite{Owens:2007kp} 
for nCTEQ15, and MSTW08~\cite{Martin:2009iq} for DSSZ12). 
The ratio Eq.~\eqref{eq:ratio} is relatively free from systematic 
uncertainties, and in particular has only a weak dependence on the input PDF 
$\tilde{f}_p$, as most dependence cancels in the ratio.

\begin{figure}[t!]
\centering
\includegraphics[width=\textwidth,clip=true,trim=3cm 0 3cm 0]{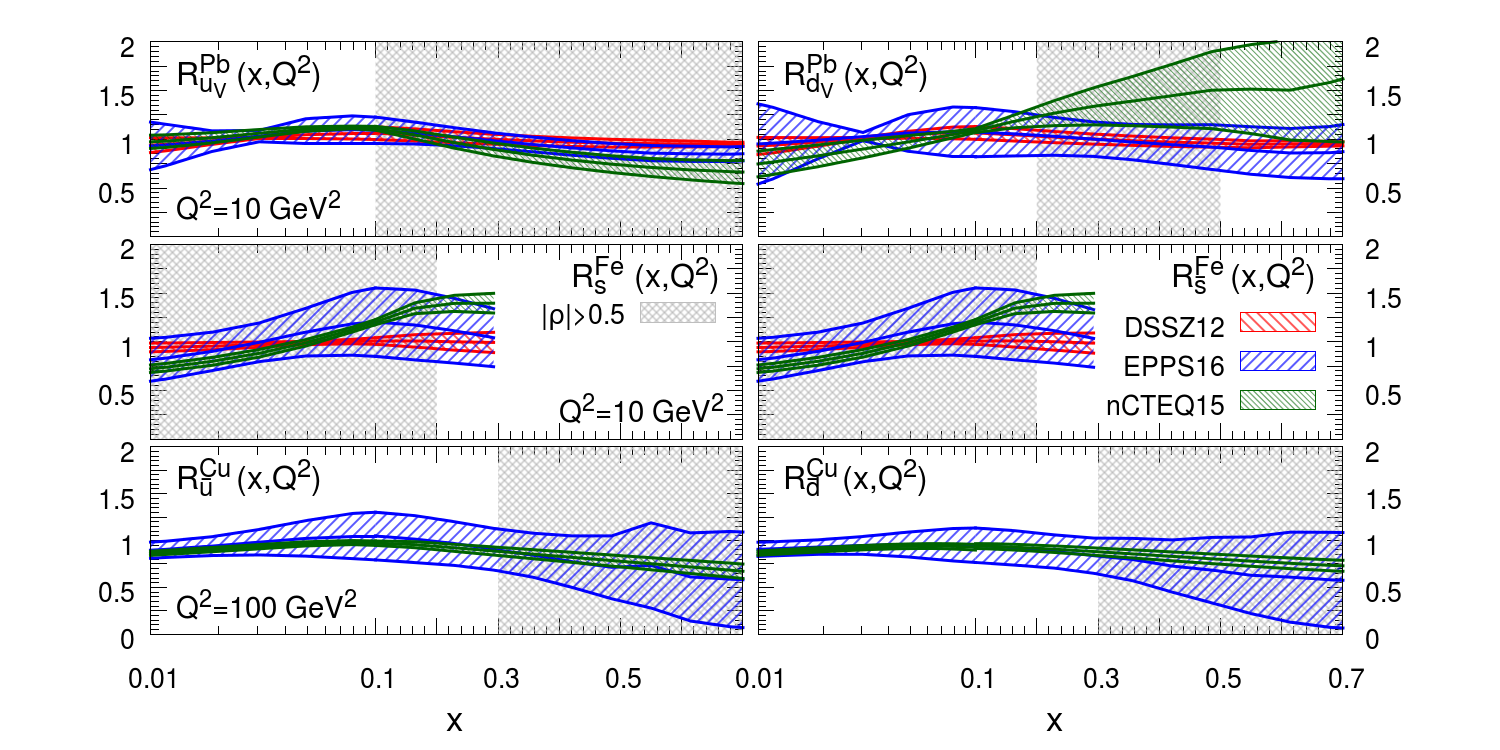}\\
\caption{The ratio $R_f^N$ between the PDFs entering the nuclear and the 
   proton observables, Eq.~\eqref{eq:ratio}, for each of the nPDF sets 
   considered in this analysis, and for the relevant flavours identified in 
   Sect.~\ref{subsec:dataset}. The shaded region corresponds to correlations
   $|\rho|>0.5$ between PDFs and observables, see also Fig.~\ref{fig:rho}.}
\label{fig:nPDFs}
\end{figure}

The observables entering Eqs.~(\ref{eq:def1})-(\ref{eq:def2b}) 
are computed~\cite{Bertone:2013vaa,zahari_kassabov_2019_2571601}
in exactly the same way for both proton and nuclear targets.
Specifically, we always take into account the non-isoscalarity of the target,
{\it i.e.}, observables are averaged over proton and neutron contributions, 
weighted by the corresponding atomic and mass numbers $Z$ and $A$ 
\begin{equation}
T^N_i[f_p] = \frac{1}{A}
\left\{
Z T_i[f_p]+(A-Z)T_i[f_n]
\right\}\,,
\qquad
T^N_i[f_N] = \frac{1}{A}
\left\{
Z T_i[f_{p/N}]+(A-Z)T_i[f_{n/N}]
\right\}\,.
\label{eq:isospin}
\end{equation}
The nuclear PDF is simply
\begin{equation}
f_N = \frac{1}{A}
\left[
Zf_{p/N}+(A-Z)f_{n/N}
\right]\,,
\label{eq:nuclpdf}
\end{equation}
and thus the only difference between proton and nuclear observables 
consists of replacing the free-proton PDF, $f_p$, with the bound-proton 
PDF, $f_{p/N}$. In all these expressions free- and bound-neutron PDFs 
are computed from their proton counterparts by assuming exact isospin symmetry.

Initially, the free-proton PDF is taken from the NNPDF3.1 set, while
the bound-proton PDF is obtained by applying to the same free-proton PDF 
the replica-by-replica nuclear correction $R_f^{N,(n)}(x,Q^2)$ determined in 
Eq.~\eqref{eq:ratio}:
\begin{equation}
f_{p/N}^{(n)} = R_f^{N,(n)} f_p\,.
\label{eq:nuccorr}
\end{equation}
In this procedure the input of the nuclear PDFs enters only through the 
correction Eq.~\eqref{eq:ratio}.
To ensure maximum theoretical consistency between proton and nuclear 
observables in Eqs.~(\ref{eq:def1})-(\ref{eq:def2b}) and \eqref{eq:isospin}, 
they are each computed with the same theoretical settings as in the fit, 
see Sect.~\ref{subsec:fitsettings}. 
The free-proton PDF $f_p$ used in Eqs.~\eqref{eq:isospin} and~\eqref{eq:nuccorr}
is then iterated to consistency, {\it i.e.}, the output of the free-proton fit 
(as described in Sect.~\ref{sec:results} below) is used to compute the
bound-proton PDF used in a subsequent fit. While in principle the PDF used in 
the determination of $R_f^{N,(n)}$ Eq.~\eqref{eq:ratio} should also be iterated, 
for consistency, in practice this is only a small correction since the 
dependence of $R_f^{N,(n)}$ on the input proton PDF is very weak.

\subsubsection{Discussion}
\label{subsubsec:discussion}

Before implementing the nuclear 
covariance matrix in a global fit of proton PDFs, we look more 
closely at the pattern of nuclear corrections, and at the 
nuisance parameters and shifts defined 
in Eqs.~(\ref{eq:def1}-\ref{eq:def2b}).
In Fig.~\ref{fig:nPDFs}, we show the nuclear correction $R_f^N$, 
Eq~\eqref{eq:ratio}.
For each nuclear species, we display only the PDF flavours that are
expected to give the largest contribution to the relevant observables,
as outlined in Sect.~\ref{subsec:dataset}.
The ratio is computed, for each of the three nPDF sets selected in the previous
section, at the typical average scale of the corresponding dataset: 
$Q^2=10$ GeV$^2$ for CHORUS and NuTeV, and $Q^2=100$ GeV$^2$ for E605.
For each nPDF set, uncertainty bands correspond to the nominal tolerances 
discussed above, and and have been rescaled to $68\%$ CLs for nCTEQ15 and 
EPPS16.
The most interesting region of maximal correlation between the PDFs and the 
observables (corresponding to $|\rho|>0.5$ in Fig.~\ref{fig:rho}) is shown 
shaded.
Since the observables depend linearly on the nuclear PDFs, the quantity
defined in Eq.~\eqref{eq:ratio} provides an estimate of the relative 
size of the nuclear corrections.
According to Fig.~\ref{fig:nPDFs}, we therefore expect moderate deviations for 
all the experiments.

\begin{figure}[!t]
\centering
\includegraphics[scale=0.23]{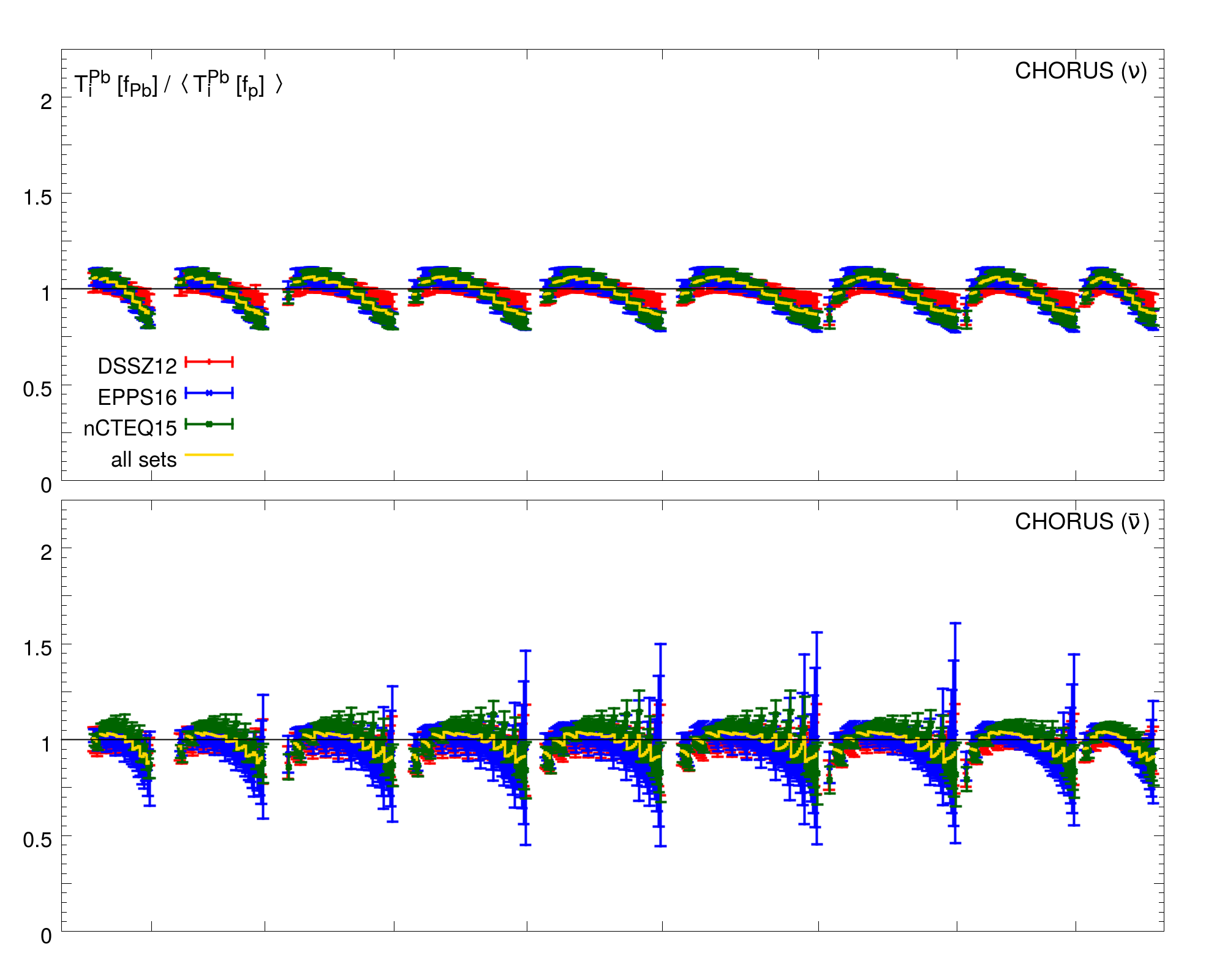}\\
\caption{The ratio between the CHORUS observable computed with nuclear PDFs,  
   $T_i^{\rm Pb}[f_{\rm Pb}]$, and the central prediction with proton PDFs,
   $\langle T_i^{\rm Pb}[f_p]\rangle$, in the case of neutrino (top) and 
   antineutrino (bottom) beams. Results from the DSSZ12, EPPS16 and nCTEQ15 
   sets are displayed separately. The ratio between the nuclear and the proton
   expectation values is also shown, averaged over the 
   combination of all replicas from all nPDF sets. In each plot there are 
   nine bins corresponding to different energies of the neutrino/antineutrino 
   beam $E_{\nu/\bar\nu}$: 25, 35, 45, 55,70, 90, 110, 130 and 170 GeV, 
   respectively. In each bin, the $x$ value increases from left to right in 
   the range displayed in Table~\ref{tab:dataset}, $0.045<x<0.65$. 
   For clarity, the various bins are separated by a tick on the 
   horizontal axis.} 
\label{fig:distancechorus}
\end{figure}
\begin{figure}[!t]
\centering
\includegraphics[scale=0.23,clip=true,trim=0 1.5cm 0 0]{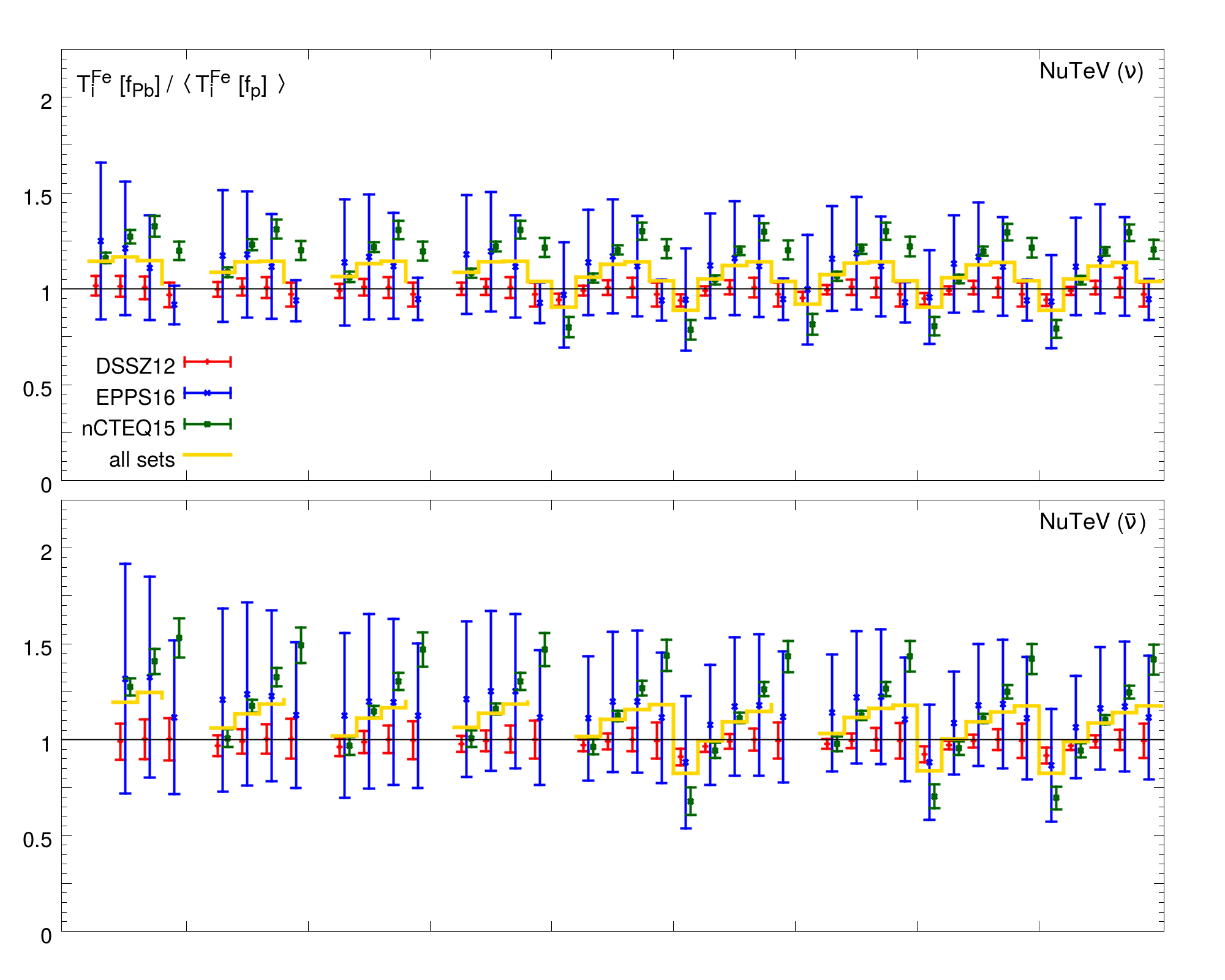}\\
\caption{Same as Fig.~\ref{fig:distancechorus}, but for NuTeV. From left to 
   right, there are nine bins corresponding to three 
   different inelasticities $y$: 0.334, 0.573, 0.790, and three different
   neutrino/antineutrino beam energies $E_{\nu/\bar\nu}$: 90.18, 174.37 and
   244.72 GeV. In each bin, the $x$ value increases from left to right in 
   the range displayed in Table~\ref{tab:dataset}, $0.02<x<0.33$ for neutrinos, 
   and $0.02<x<0.21$ for antineutrinos.}
\label{fig:distancenutev}
\end{figure}
\begin{figure}[!t]
\centering
\includegraphics[scale=0.23]{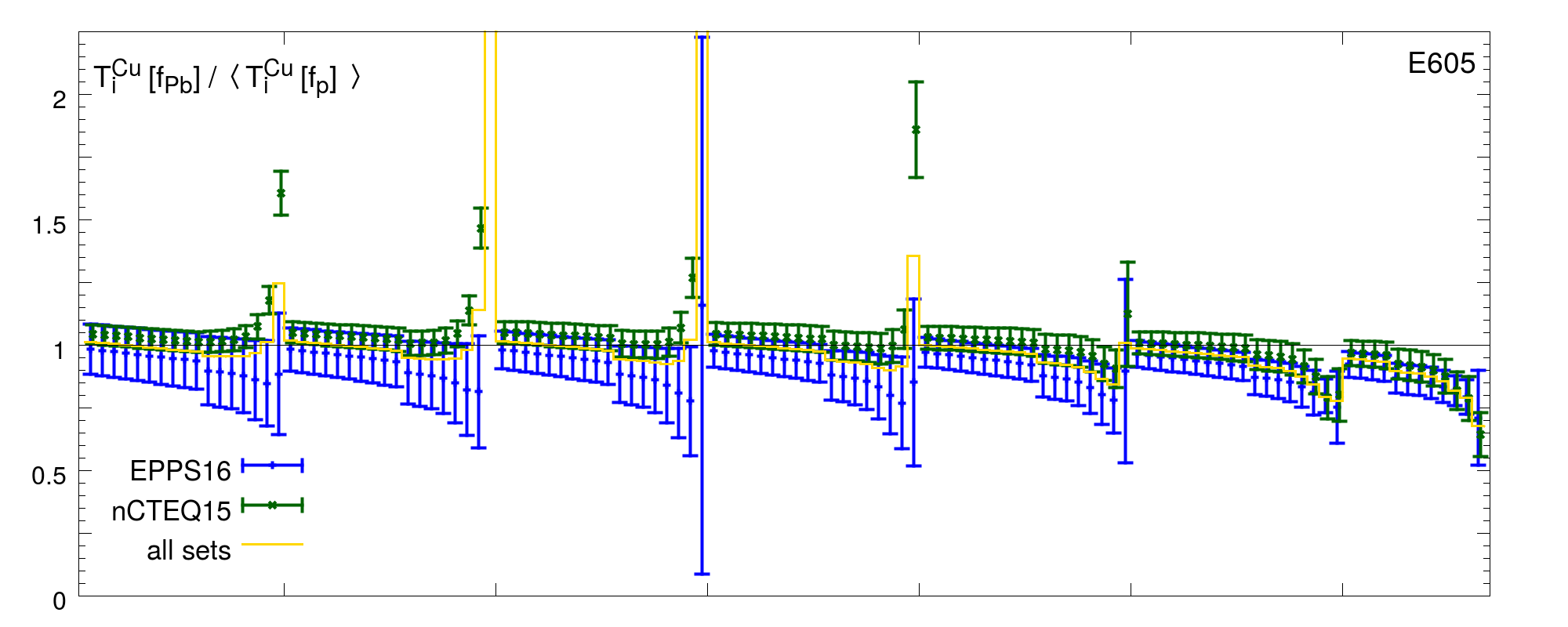}\\
\caption{Same as Fig.~\ref{fig:distancechorus}, but for E605. From left to 
   right, there are seven bins corresponding to different
   values of the vector boson rapidity $y$: -0.2, -0.1, 0.0, 0.1, 0.2, 0.3 and
   0.4. In each bin, the value of the dilepton invariant mass $M_{\ell\ell}$ 
   increases from left to right in the range displayed in 
   Table~\ref{tab:dataset}, $7.1<M_{\ell\ell}<10.9$ GeV.}
\label{fig:distancedye605}
\end{figure}

The nuclear corrections to the observables themselves are shown in 
Figs.~\ref{fig:distancechorus}-\ref{fig:distancedye605}, where, for each data 
point $i$ (after kinematic cuts) in the CHORUS, NuTeV and E605 datasets, we 
display the observables computed with nuclear PDFs, normalised to the 
expectation value with the proton PDF, $T_i^N[f_N]/\langle T_i^N[f_p]\rangle$.
Results are shown for the DSSZ12, EPPS16 and nCTEQ15 sets separately.
The central value of the same ratio obtained from the 
Monte Carlo combination of all the three nPDF sets is also shown. 
Looking at Figs.~\ref{fig:distancechorus}-\ref{fig:distancedye605}, we observe 
that the shape and size of the ratios between observables closely follow 
that of the ratios between PDFs in the shaded region of Fig.~\ref{fig:nPDFs}.
For CHORUS, all the nPDFs modify the nuclear observable in a similar way,
with a slight enhancement at smaller values of $x$, and a suppression at 
larger values of $x$. 
Overall, the uncertainty in the nuclear observable is comparable 
for the three sets. 
All sets are mutually consistent within uncertainties.
For NuTeV, the nuclear corrections are markedly inconsistent for DSSZ12 and 
nCTEQ15, with the former being basically flat around one, while the latter
deviates very significantly from one in some bins.
Both nPDF sets are included within the much larger uncertainties of 
the EPPS16 set, which has now the largest stated uncertainty, 
especially in the case of antineutrino beams.
This is likely a consequence of the fact that the strange quark nPDF was
fitted independently from the other quark nPDFs in EPPS16, while it was
related to lighter quark nPDFs in DSS12 and nCTEQ15~\cite{Paukkunen:2018kmm}.
For E605, the nuclear correction gives a mild reduction in most of   
the measured kinematic range.
The nCTEQ15 nPDF set appears to be more precise than the EPPS16 set, and 
they are reasonably consistent with each other. 
No DSSZ12 set is available for Cu.
On average the size of the nuclear correction shift, as quantified by the ratio
$T_i^N[f_N]/T_i^N[f_p]$, is of order $10\%$ for CHORUS, $20\%$ for NuTeV, and 
$5\%$ for E605. 

\begin{figure}[!t]
\centering
\includegraphics[scale=0.294,clip=true,trim=1cm 3cm 0 1.5cm]{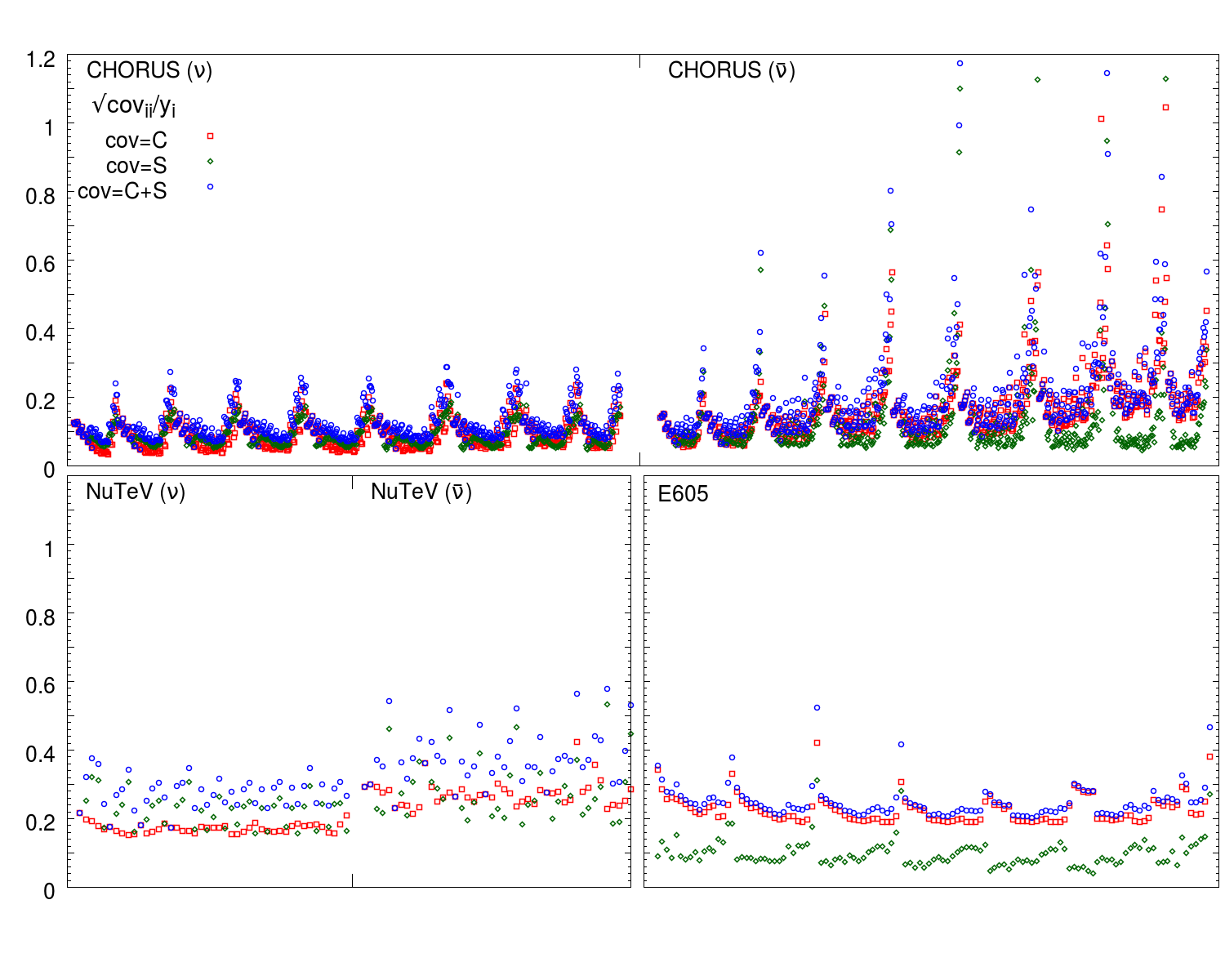}\\
\caption{The square root of the diagonal elements of the covariance matrix 
   normalised to the experimental data, 
   $\sqrt{{\rm cov}_{ii}}/D_i$, for CHORUS, NuTeV and E605. 
   We show results for the experimental covariance 
   matrix, $C_{ij}$, for the nuclear covariance matrix, $S_{ij}$, computed from 
   Eq.~\eqref{eq:def1}, and for their sum.}
\label{fig:diag}
\end{figure}

To gain a further idea of the effects to be expected from the nuclear 
corrections, in Fig.~\ref{fig:diag} we show the 
square root of the diagonal elements of the experimental and theoretical 
covariance matrices, and their sum, each normalised
to the central value of the experimental data: $\sqrt{{\rm cov}_{ii}}/D_i$
(where ${\rm cov}_{ii}$ is equivalent, respectively, to $C_{ii}$, $S_{ii}$ or
$C_{ii}+S_{ii}$).
The theoretical covariance matrix is computed with Eq.~\eqref{eq:nuisance} 
using the nuisance parameters Eq.~\eqref{eq:def1};
the general pattern of the results does not change qualitatively if 
Eqs.~\eqref{eq:def2a}-\eqref{eq:def2b} are used instead.
The features observed in Figs.~\ref{fig:distancechorus}-\ref{fig:distancedye605}
are paralleled in Fig.~\eqref{fig:diag}; in particular, the
dependence of the size of the nuclear uncertainties on the bin kinematics.
Moreover we can now see that the nuclear uncertainties are much smaller than 
the data uncertainties for E605, while for CHORUS (particularly $\nu$) they can 
be comparable. However for NuTeV the nuclear uncertainties are rather 
larger than the experimental uncertainties.
This suggests that the NuTeV data will have relatively less weight in 
the global fit once the nuclear uncertainties are accounted for.

\begin{figure}[!t]
\centering
\includegraphics[scale=0.55,clip=true,trim=0.8cm 0 1.2cm 0]{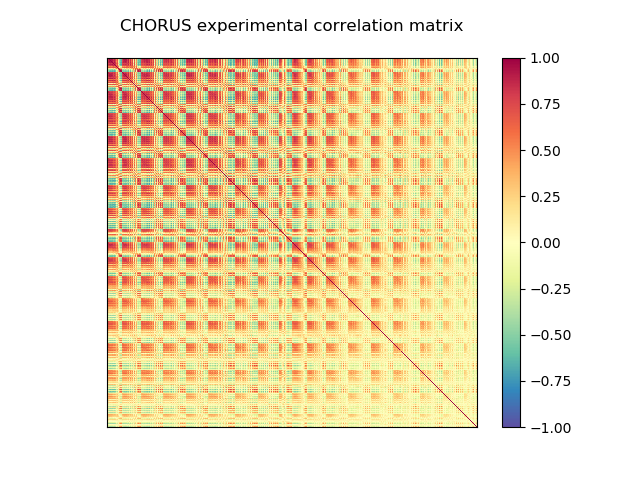}
\includegraphics[scale=0.55,clip=true,trim=0.8cm 0 1.2cm 0]{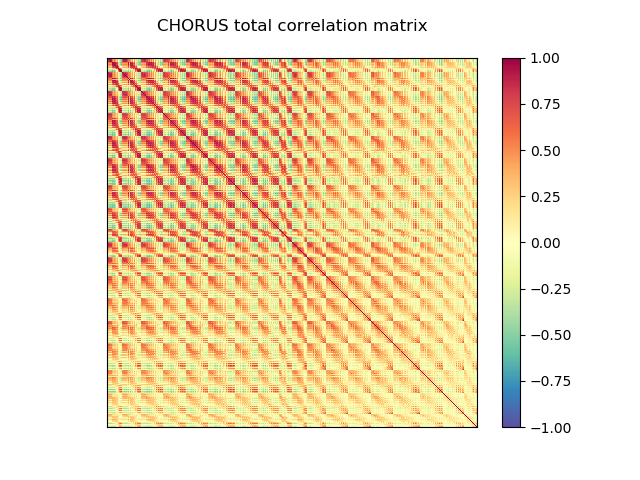}\\
\includegraphics[scale=0.55,clip=true,trim=0.8cm 0 1.2cm 0]{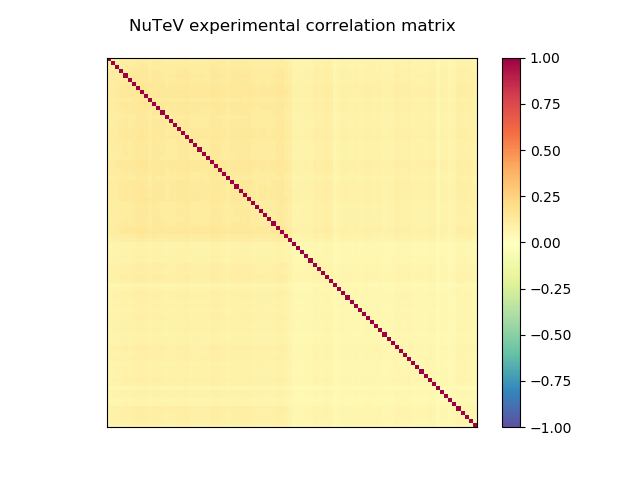}
\includegraphics[scale=0.55,clip=true,trim=0.8cm 0 1.2cm 0]{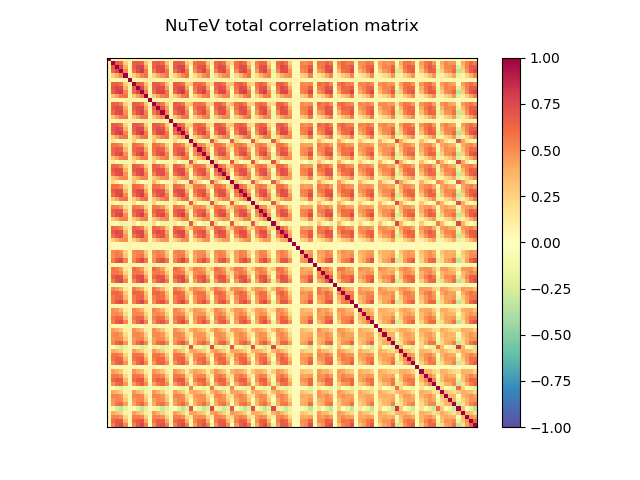}\\
\includegraphics[scale=0.55,clip=true,trim=0.8cm 0 1.2cm 0]{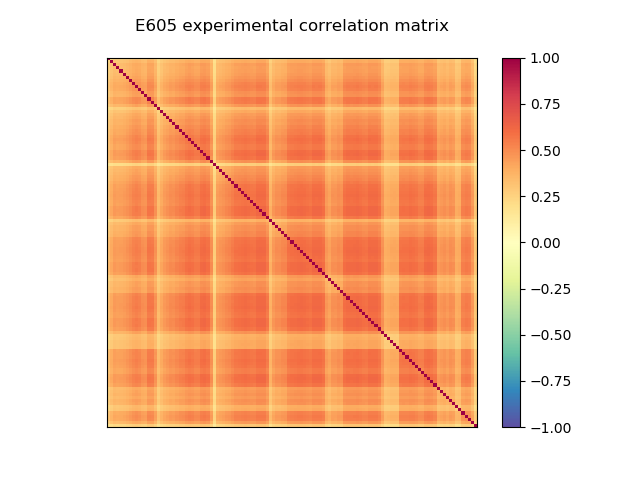}
\includegraphics[scale=0.55,clip=true,trim=0.8cm 0 1.2cm 0]{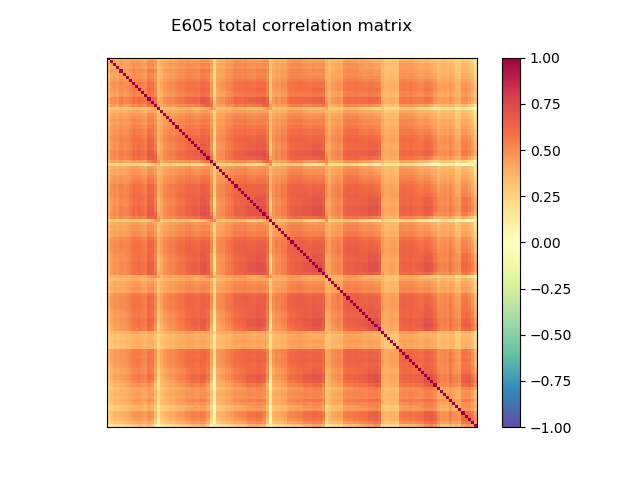}\\
\caption{The experimental (left) and total (right) correlation matrices 
   for CHORUS, NuTeV and E605. The theoretical covariance matrix, added to
   the experimental covariance matrix to obtain the total covariance matrix,  
   is computed according to Eq.~\eqref{eq:def1}.}
\label{fig:corrplot}
\end{figure}

Finally, in Fig.~\ref{fig:corrplot} we show the experimental correlation 
matrices $\rho^C_{ij} = C_{ij}/\sqrt{C_{ii}C_{jj}}$ as heat plots: correlated 
points are red, while anti-correlated points are blue. For comparison we 
show the total correlation
matrix obtained by summing the experimental and theoretical covariance
matrices: $\rho^{C+S}_{ij} = (C_{ij}+S_{ij})/\sqrt{(C_{ii}+S_{ii})(C_{jj}+S_{jj})}$.
These are computed according to Eq.~\eqref{eq:def1} but the qualitative
behaviour of the total correlation matrix is unaltered if
Eqs.~\eqref{eq:def2a}-\eqref{eq:def2b} are used instead.
We see that our procedure captures the sizeable correlations of 
the nuclear corrections between different bins of momentum and energy,
systematically enhancing bin-by-bin correlations in the data.
Note as we might expect the nuclear uncertainties are strongly correlated 
between the different sets in the same experiment ({\it i.e.}, neutrino 
and antineutrino sets in CHORUS and NuTeV).
In principle, predictions for data points belonging to experiments that 
use different nuclear targets should be somewhat correlated, because 
part of the fit parameters in the nPDF analyses control the dependence 
on the nuclear mass number $A$.
In order to be conservative, however, we do not attempt to include 
these correlations of the nuclear uncertainties between the experiments 
on different nuclear targets.
If information  were reliably included, it would
reduce the overall effect of the nuclear uncertainty.

\vfill\eject

\newpage
\newpage
\section{Impact of Theoretical Corrections in a Global PDF Fit}
\label{sec:results}

In this section, we discuss the impact of the theoretical uncertainties due 
to nuclear corrections, as computed in the previous section,  
in a global fit of proton PDFs.
We first summarise the experimental and theoretical settings of the fits,
then we present the results. 

\subsection{Fit Settings}
\label{subsec:fitsettings}

The PDF sets discussed in this section are based on a variant of the 
NNLO NNPDF3.1 global analysis~\cite{Ball:2017nwa}.
In particular the experimental input, and
related kinematic cuts, are exactly the same as in the NNPDF3.1 NNLO fit.
On top of the nuclear measurements presented in 
Sect.~\ref{subsec:dataset}, the dataset is made up of: 
fixed-target~\cite{Arneodo:1996kd,Arneodo:1996qe,Benvenuti:1989rh,
Benvenuti:1989fm,Whitlow:1991uw,Onengut:2005kv,Goncharov:2001qe,Mason:2006qa} 
and collider~\cite{Abramowicz:2015mha} DIS inclusive structure
functions; charm and botton cross sections from HERA~\cite{Abramowicz:1900rp};
fixed-target DY cross sections~\cite{Webb:2003ps,Webb:2003bj,Towell:2001nh};
gauge boson and inclusive jet production cross sections from the 
Tevatron~\cite{Aaltonen:2010zza,Abazov:2007jy,Aaltonen:2008eq,Abazov:2013rja,
D0:2014kma}; and electroweak boson production, inclusive jet, $Z$ $p_T$, total 
and differential top-pair cross sections from ATLAS~\cite{Aad:2011dm,
Aad:2013iua,Aad:2011fp,Aad:2011fc,Aad:2013lpa,ATLAS:2012aa,ATLAS:2011xha,
TheATLAScollaboration:2013dja,Aad:2015auj,Aaboud:2016btc,Aad:2014kva,
Aaboud:2016pbd,Aad:2015mbv,Aad:2014qja,Aad:2014xaa},
CMS~\cite{Chatrchyan:2012xt,Chatrchyan:2013mza,Chatrchyan:2013tia,
Chatrchyan:2013uja,Chatrchyan:2013faa,Chatrchyan:2012bra,Chatrchyan:2012ria,
Khachatryan:2016pev,Khachatryan:2015luy,Khachatryan:2016mqs,
Khachatryan:2015oqa,Khachatryan:2015oaa} and LHCb~\cite{Aaij:2012vn,
Aaij:2012mda,Chatrchyan:2012bja,Aaij:2015gna,Aaij:2015zlq}.
The theoretical input is also the same as in NNPDF3.1: the strong running 
coupling at the $Z$-boson mass is fixed to $\alpha_s(m_Z)=0.118$, 
consistent with the PDG average~\cite{Patrignani:2016xqp}; heavy-quark mass 
effects are included using the FONLL C general-mass 
scheme~\cite{Forte:2010ta,Ball:2015tna}, with pole masses $m_c=1.51$ GeV for
charm and $m_b=4.92$ GeV for bottom, consistent with the Higgs cross section
working group recommendation~\cite{deFlorian:2016spz}; the charm PDF is fitted 
in the same way as the other light quark PDFs~\cite{Ball:2016neh};
and the initial parametrisation scale is chosen just above
the value of the charm mass, $Q_0=1.65$ GeV.
All fits are performed at NNLO in pure QCD, and result in ensembles
of $N_{\rm rep}=100$ replicas.

In comparison to NNPDF3.1, we have made small improvements in the 
computation of the  CHORUS and NuTeV observables.
In the case of CHORUS, cross sections were computed in NNPDF3.1 following the 
original implementation of Ref.~\cite{Ball:2008by}, where the target was 
assumed to be isoscalar, and the data were supplemented with a systematic 
uncertainty to account for their actual non-isoscalarity.
We now remove this uncertainty, and we compute the cross sections taking into
account the non-isoscalarity of the target, as explained in 
Sect.~\ref{subsubsec:implementation}. This increases the $\chi^2$ per 
data point a little (from $1.11$ to $1.25$). 
In the case of NuTeV, we update the value of the branching ratio of charmed 
hadrons into muons.
This value, which is used to reconstruct charm production 
cross sections from the neutrino dimuon production cross sections measured by 
NuTeV, was set equal to $0.099$ in NNPDF3.1, following the original
analysis of Ref.~\cite{Ball:2009mk}.
Previously, the uncertainty on the branching ratio was not taken into account.
We now utilise the current PDG result, 
$0.086\pm 0.005$~\cite{Patrignani:2016xqp}, and include its uncertainty 
as an additional fully correlated systematic uncertainty. This reduces 
the $\chi^2$ per data point a little (from $0.82$ to $0.66$).

With these settings, we perform the following four fits: 
\begin{itemize}
\item a {\it Baseline} fit,
based on the theoretical and experimental inputs described above, and without
any inclusion of theoretical nuclear uncertainties; 
\item a ``No Nuclear'' fit, {\it NoNuc}, equal to the Baseline, but without the datasets 
that utilise nuclear targets, {\it i.e.} without CHORUS, NuTeV and E605; 
\item a ``Nuclear Uncertainties'' fit, {\it NucUnc}, equal to the Baseline, but with the 
inclusion of theoretical nuclear uncertainties applied to CHORUS, NuTeV 
and E605, according to 
Eqs.~\eqref{eq:datagenth}-\eqref{eq:chi2th}, the theory covariance matrix 
being computed using \eqref{eq:nuisance} with the prescription 
Eq.~\eqref{eq:def1} for the nuisance paremeters;
\item a ``Nuclear Corrections'' fit, {\it NucCor}, equal to the 
Baseline, but with the 
inclusion of theoretical nuclear uncertainties applied to CHORUS, NuTeV 
and E605, according to 
Eqs.~\eqref{eq:datagenth}-\eqref{eq:chi2th}, with the nuclear 
correction $\delta T_i^N$, Eq.~\eqref{eq:def2b}, added to the 
theoretical prediction 
$T_i[f]$ used in Eq.~\eqref{eq:chi2th}, and the theory covariance matrix 
computed using Eq.~\eqref{eq:nuisance} and Eq.~\eqref{eq:def2a} 
for the nuisance parmeters.
\end{itemize} 
The results of these fits are presented below.

\subsection{Fit Quality and Parton Distributions}
\label{subsec:fits}

We first discuss the quality of the fits. 
In Table~\ref{tab:chi2} we report the values of the $\chi^2$ per data
point for each of the four fits listed above.
Values are displayed for separate datasets, or groups of datasets 
corresponding to measurements of similar observables in the 
same experiment, and for the total dataset.
We take into account all correlations in the computation of the experimental
and/or theoretical covariance matrices entering the definition of the $\chi^2$,
Eq.~\eqref{eq:chi2th}; multiplicative uncertainties are treated using the $t_0$ 
method~\cite{Ball:2009qv}, and two fit iterations are performed to ensure 
convergence of the final results.

\begin{table}[!t]
\centering
\scriptsize
\renewcommand{\arraystretch}{1.13}
\begin{tabularx}{\textwidth}{lcCCCC}
\toprule
  Experiment
& $N_{\rm dat}$
& Baseline 
& NoNuc
& NucUnc 
& NucCor\\
\midrule
NMC                      
&  325 & 1.31 & 1.33 & 1.31 & 1.29 \\
SLAC                     
&   67 & 0.79 & 0.93 & 0.72 & 0.73 \\
BCDMS                    
&  581 & 1.23 & 1.17 & 1.20 & 1.21 \\
\bf CHORUS $\nu$             
&  \bf 416 & \bf 1.29 & ---  & \bf 0.97 & \bf 1.04 \\
\bf CHORUS $\bar{\nu}$       
&  \bf 416 & \bf 1.20 & ---  & \bf 0.78 & \bf 0.83 \\
\bf NuTeV dimuon $\nu$       
&   \bf 39 & \bf 0.41 & ---  & \bf 0.31 & \bf 0.40 \\
\bf NuTeV dimuon $\bar{\nu}$ 
&   \bf 37 & \bf 0.90 & ---  & \bf 0.62 & \bf 0.83 \\
\midrule
HERA I+II incl.          
& 1145 & 1.15 & 1.15 & 1.15 & 1.15 \\
HERA $\sigma_c^{\rm NC}$   
&   37 & 1.40 & 1.52 & 1.46 & 1.44 \\
HERA $F_2^b$             
&   29 & 1.11 & 1.10 & 1.10 & 1.11 \\
\midrule
E866 $\sigma^d_{\rm DY}/\sigma^p_{\rm DY}$    
&   15 & 0.47 & 0.33 & 0.44 & 0.44 \\
E886 $\sigma^p$                             
&   89 & 1.35 & 1.22 & 1.69 & 1.66 \\
\bf E605  $\sigma^p$                            
&  \bf 85  & \bf 1.18 & ---  & \bf 0.85 & {\bf 0.89} \\
CDF $Z$ rap                                    
&   29 & 1.41 & 1.29 & 1.39  & 1.41 \\
CDF Run II $k_t$ jets                          
&   76 & 0.88 & 0.82 & 0.89 & 0.92 \\
D0 $Z$ rap                                     
&   28  & 0.60 & 0.57 & 0.59 & 0.59 \\
D0 $W$  asy                            
&    17  & 2.11 & 2.06 & 2.10 & 2.09 \\
\midrule
ATLAS total                                
&  360  & 1.08 & 1.04 & 1.04 & 1.05 \\
ATLAS $W,Z$ 7 TeV 2010                     
&   30  & 0.93 & 0.93 & 0.92 & 0.93 \\
ATLAS HM DY 7 TeV                          
&    5  & 1.68 & 1.60 & 1.57 & 1.54 \\
ATLAS low-mass DY 7 TeV                     
&    6 & 0.91 & 0.87 & 0.88 & 0.89 \\
ATLAS $W,Z$ 7 TeV 2011                     
&   34 & 1.97 & 1.78 & 1.87 & 1.94 \\
ATLAS jets                       
&   180 & 1.00 & 0.97 & 1.00 & 1.00 \\
ATLAS $Z$ $p_T$ 8 TeV   
&   92 & 0.93 & 0.95 & 0.95 & 0.92 \\
ATLAS $t\bar{t}$                      
&   13 & 1.32 & 1.22 & 1.24  & 1.21 \\
\midrule
CMS total                              
&  409  & 1.07 & 1.07 & 1.07 & 1.07 \\
CMS W asy                   
&   22 & 1.23 & 1.41 & 1.30 & 1.31 \\
CMS Drell-Yan 2D 2011                  
&  110 & 1.30 & 1.33 & 1.31 & 1.29 \\
CMS W rap 8 TeV                     
&   22 & 0.94 & 0.88 & 0.95 & 0.96 \\
CMS jets                 
&  214 & 0.93 & 0.89 & 0.90 & 0.91 \\
CMS $Z$ $p_T$ 8 TeV $(p_T^{ll},y_{ll})$ 
&   28 & 1.31 & 1.38 & 1.35 & 1.35 \\
CMS $t\bar{t}$                    
&   13  & 0.77 & 0.73 & 0.76 & 0.75 \\
\midrule
LHCb total                              
&   85 & 1.46 & 1.27 & 1.32 & 1.37 \\
LHCb $Z$                            
&    26  & 1.25 & 1.25 & 1.26 & 1.25 \\
LHCb $W,Z\to \mu$                  
&   59 & 1.55 & 1.28 & 1.35  & 1.42 \\
\midrule
{\bf Total }
& {\bf 4285}   
& {\bf 1.177}  
& {\bf 1.144} 
& {\bf 1.073}
& {\bf 1.086}\\
\bottomrule\\
\end{tabularx}
\vspace{0.5cm}
\caption{\small The values of the $\chi^2$ per data point for the various fits
  described in the text. Data sets that utilise a nuclear target other than 
  a deuteron are highlighted in boldface.}
\label{tab:chi2}
\end{table}


Inspection of Table~\ref{tab:chi2} reveals that the global fit quality
improves either if nuclear data are removed, or if they are retained
with the supplemental theoretical uncertainty.
The lowest global $\chi^2$ is obtained when the theoretical covariance
matrix is included, with the deweighted implementation NucUnc leading 
to a slightly lower value than the corrected implementation NucCor. 
In all cases, the improvent is mostly driven by the fact that the $\chi^2$ 
for all the Tevatron and LHC hadron collider experiments decreases.
This suggests that there might be some tension between nuclear 
and hadron collider data in the global fit.

Interestingly, however, the relatively poor $\chi^2$ of the ATLAS
$W,Z$ 7 TeV 2011 dataset, which is sensitive to the strange PDF,
only improves a little if the NuTeV dataset, which is also 
sensitive to the strange PDF, is either removed from the fit or 
supplemented with the theoretical uncertainty.
This suggests that the poor $\chi^2$ of the ATLAS $W,Z$ 7 TeV 
2011 dataset does not arise from tension with the NuTeV dataset.
The two datasets are indeed sensitive to different kinematic regions, and 
there is little interplay between the two, as we will further demonstrate in 
Sect.~\ref{subsec:strange}. 

The fit quality of the DIS and fixed-target DY data is stable across the fits.
Fluctuations in the values of the corresponding $\chi^2$ are small, except 
for a slight worsening in the $\chi^2$ of the HERA charm cross 
sections and of the fixed-target proton DY cross section.
Concerning nuclear datasets, their $\chi^2$ always decreases in the NucUnc and
NucCor fits in comparison to the Baseline fit, with the size of the decrease
being slightly larger in the NucUnc fit than in the NucCor fit.
This suggests that when the shift is used as a nuclear 
correction, such a correction is reasonably reliable, in the sense that its 
uncertainty is not substantially underestimated. However, the 
NucUnc implementation is more conservative than the NucCor one, and leads 
to a better fit.

\begin{figure}[!t]
\centering
\includegraphics[scale=0.47]{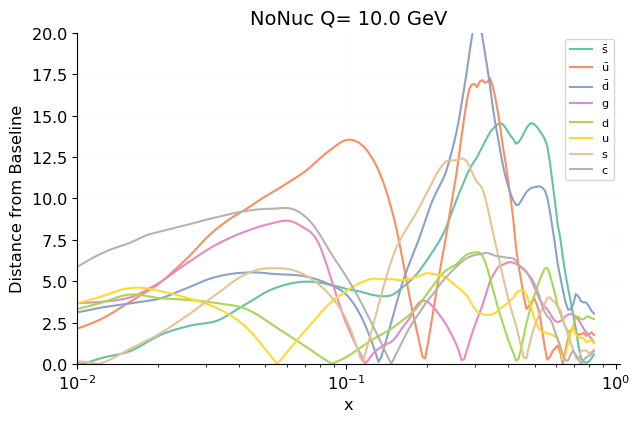}\ \
\includegraphics[scale=0.47]{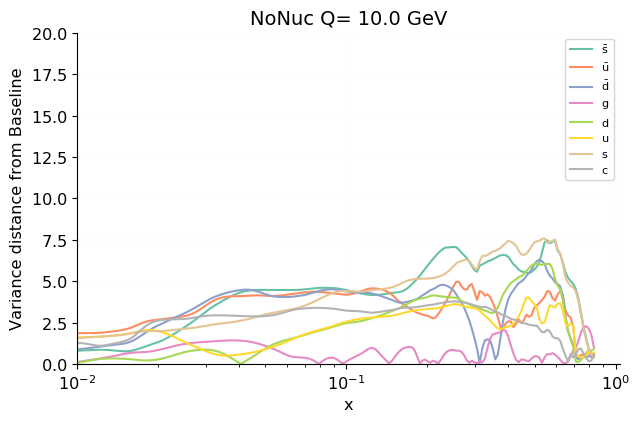}\\
\includegraphics[scale=0.47]{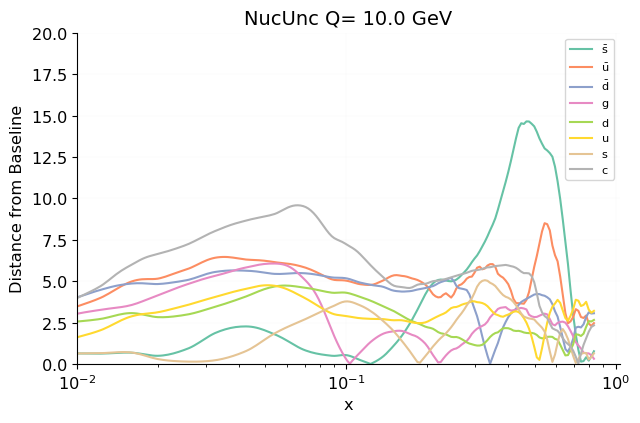}\ \
\includegraphics[scale=0.47]{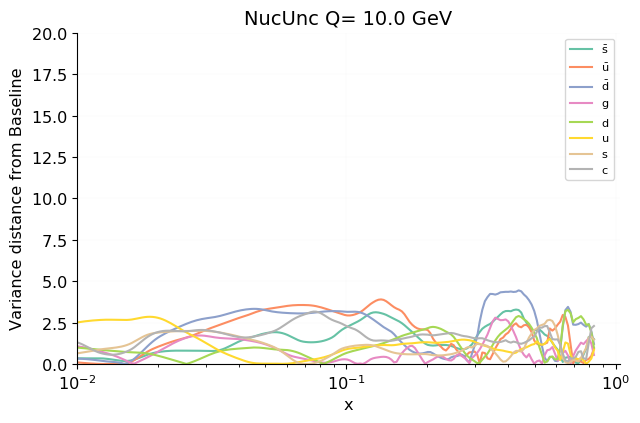}\\
\includegraphics[scale=0.47]{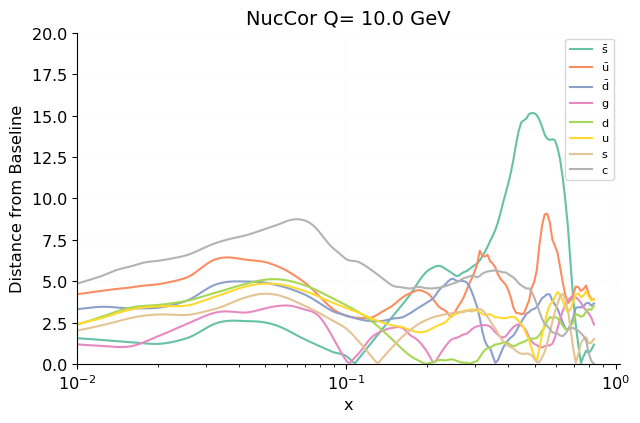}\ \
\includegraphics[scale=0.47]{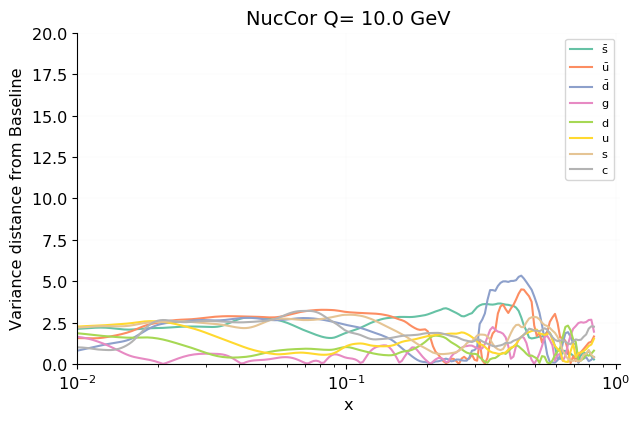}\\
\caption{Distances between the central values (left) and the uncertainties 
  (right) of the Baseline fit and the NoNuc (top), NucUnc 
  (middle), and NucCor (bottom) fits, see text for details. Results
  are displayed as a function of $x$ at a representative scale of the 
  nuclear dataset, $Q=10$ GeV, for all PDF flavours.}
\label{fig:distances}
\end{figure}

We now compare the PDFs obtained from the various fits, and study how their
central values and uncertainties vary.
To do so, we first inspect the distance between the Baseline and 
each of the other fits.
The distance between two fits, defined, {\it e.g.}, in Ref~\cite{Ball:2010de},
quantifies their statistical equivalence.
Specifically, for two PDF sets made of $N_{\rm rep}=100$ replicas, a
distance of $d\simeq 1$ corresponds to statistically equivalent sets,
while a distance of $d\simeq 10$ corresponds to sets that differ by 
one-sigma in units of the corresponding standard deviation.
We display the distance between each pair of fits, both for the central value 
and for the uncertainty, in Fig.~\ref{fig:distances}.
Results are displayed as a function of $x$ at a representative scale of the 
nuclear dataset, $Q=10$ GeV, for all PDF flavours.

As expected, the largest distances with respect to the Baseline fit are 
displayed by sea quarks, at the level of both the central value and the 
uncertainty, to which the nuclear dataset is mostly sensitive.
In the NoNuc fit, the central values of the $\bar{u}$, $\bar{d}$, $s$ and 
$\bar{s}$ quarks can differ by more than one sigma, while the 
distance in the corresponding uncertainties is more limited, albeit 
still around half a sigma, especially for $s$ and $\bar{s}$ PDFs.
This confirms that nuclear data still have a sizeable weight in a global 
fit, as already noted in Sect.~4.11 of Ref.~\cite{Ball:2017nwa}.
In the global fits including theoretical uncertainties from nuclear 
effects, the pattern of the 
distances to the Baseline is mostly insensitive to whether the 
nuclear effects are considered as an uncertainty or a correction.
For the central value, distances are always below one sigma, except for the
$\bar{s}$ PDF, where differences with respect to
the Baseline fit exceed one sigma in the range $0.4\leq x\leq 0.6$.
For the uncertainties, distances from the Baseline are comparable in the two 
fits NucUnc and NucCor, with $\bar{u}$, $\bar{d}$ $s$ and $\bar{s}$ PDFs
displaying the largest values, which are however all no more than half a sigma.

\begin{figure}[!t]
\centering
\includegraphics[scale=0.47]{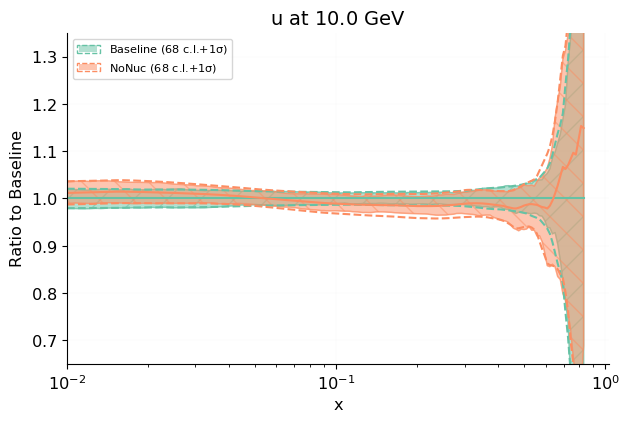}
\includegraphics[scale=0.47]{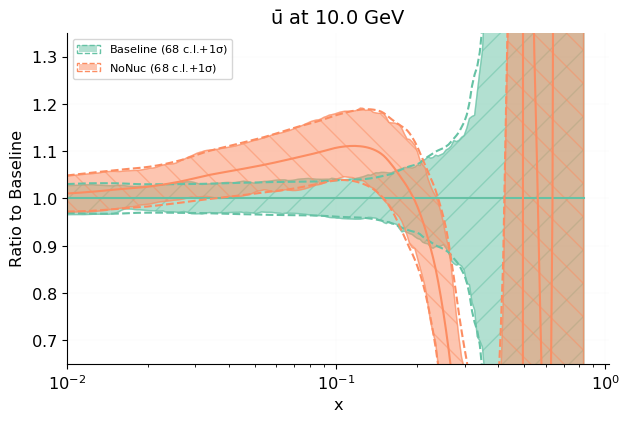}\\
\includegraphics[scale=0.47]{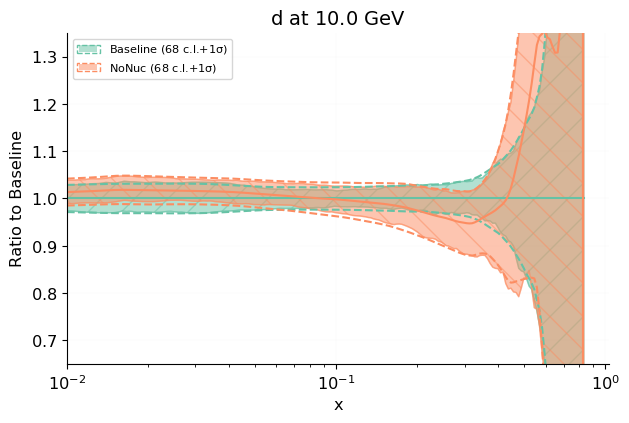}
\includegraphics[scale=0.47]{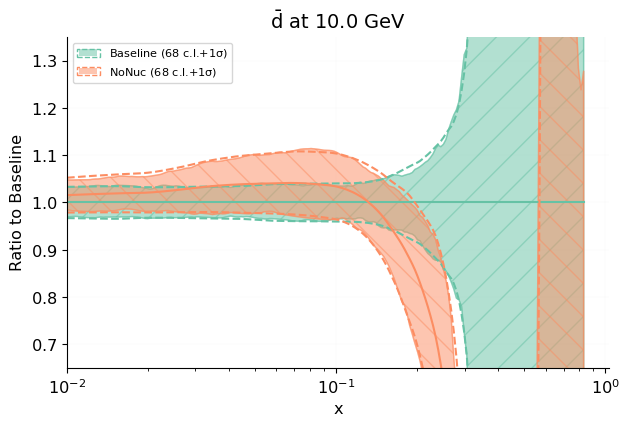}\\
\includegraphics[scale=0.47]{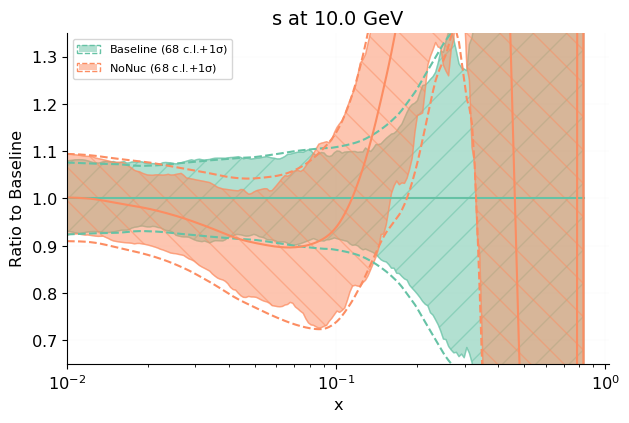}
\includegraphics[scale=0.47]{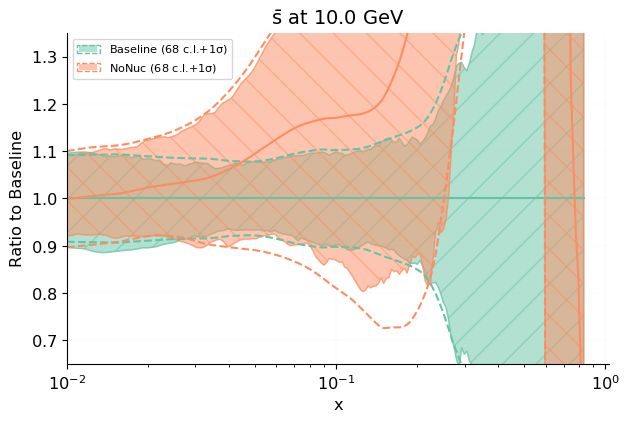}\\
\caption{Comparison between the Baseline fit, and the fit without any data on
   nuclear targets, NoNuc. The light quark (left) and antiquark
   (right) flavours are shown at $Q=10$ GeV. Results are normalised to the 
   Baseline fit.}
\label{fig:xpdf_cv_nonucl}
\end{figure}
\begin{figure}[!t]
\centering
\includegraphics[scale=0.47]{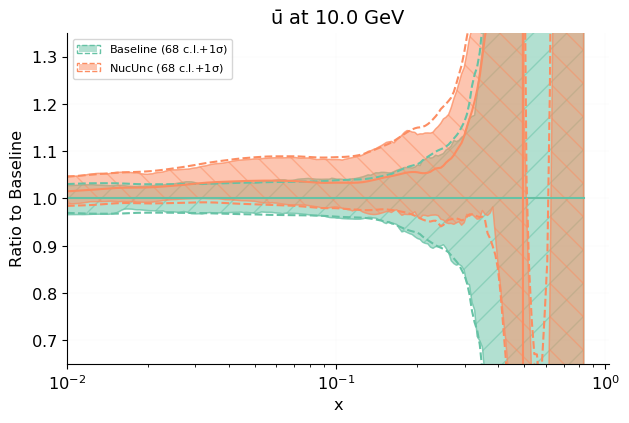}
\includegraphics[scale=0.47]{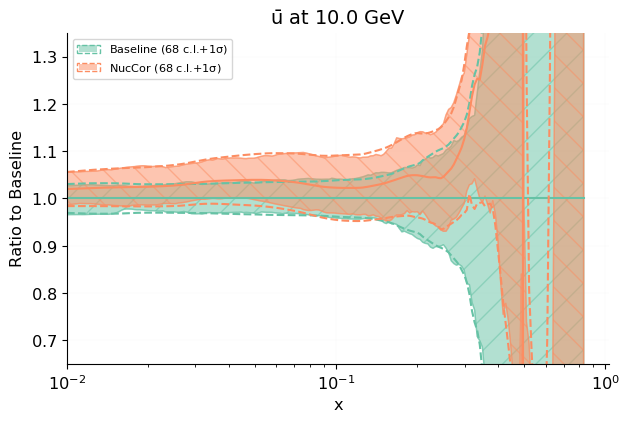}\\
\includegraphics[scale=0.47]{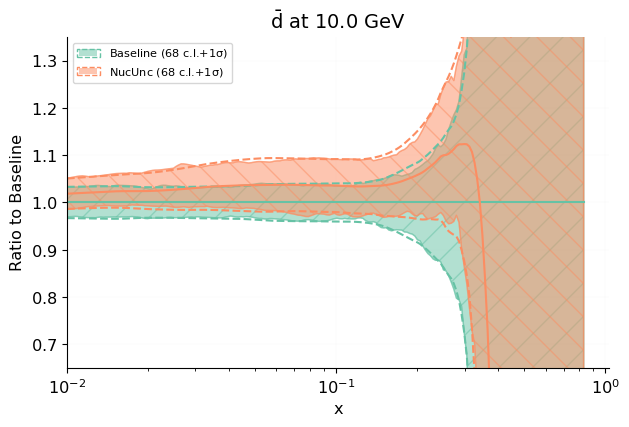}
\includegraphics[scale=0.47]{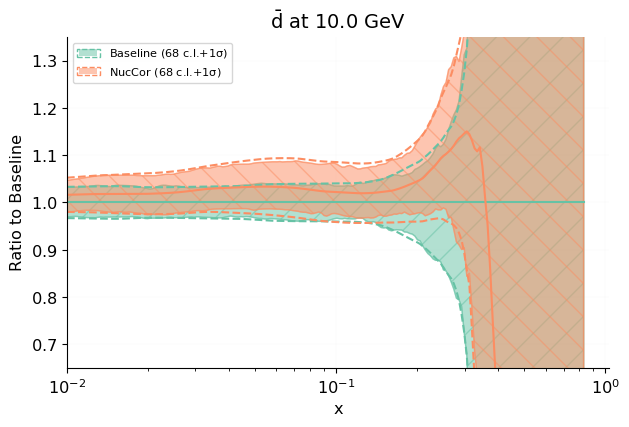}\\
\includegraphics[scale=0.47]{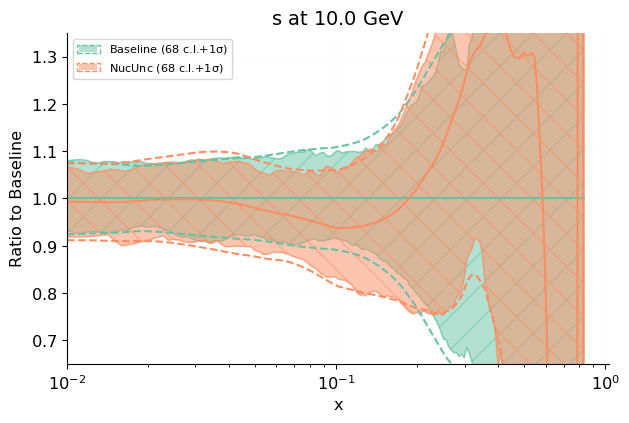}
\includegraphics[scale=0.47]{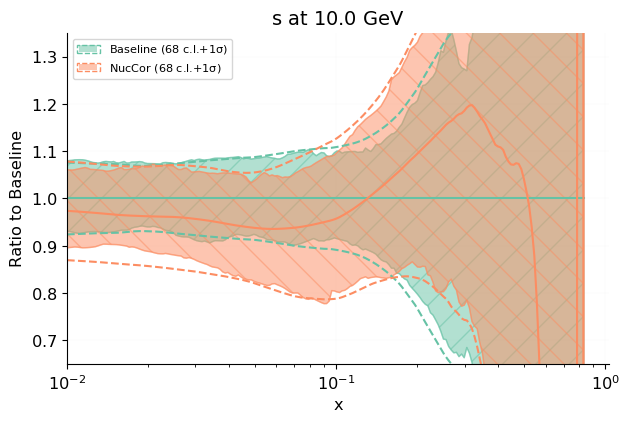}\\
\includegraphics[scale=0.47]{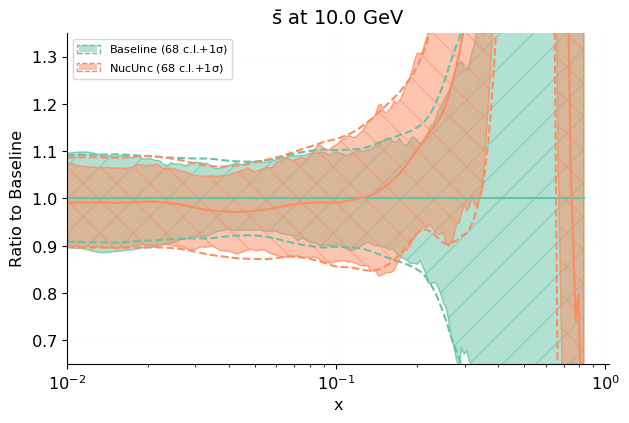}
\includegraphics[scale=0.47]{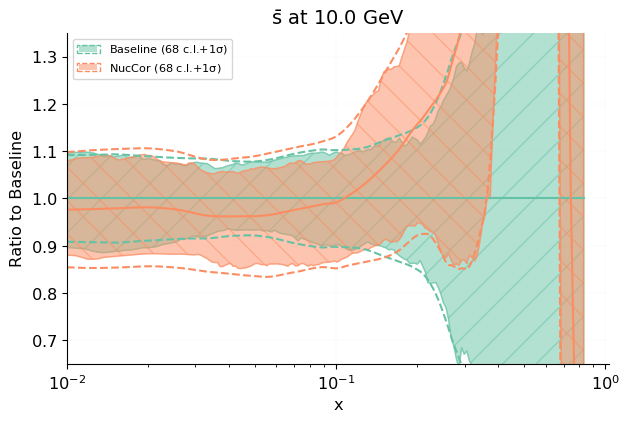}\\
\caption{Comparison between the Baseline fit, and each fit including theoretical
   uncertainties, NucUnc (left), and NucCor (right). The  
   antiup, antidown, strange and antistrange flavours are shown at $Q=10$ GeV
   from top to bottom. Results are normalised to the Baseline fit.}
\label{fig:xpdf_cv_nucl}
\end{figure}
\begin{figure}[!t]
\centering
\includegraphics[scale=0.47]{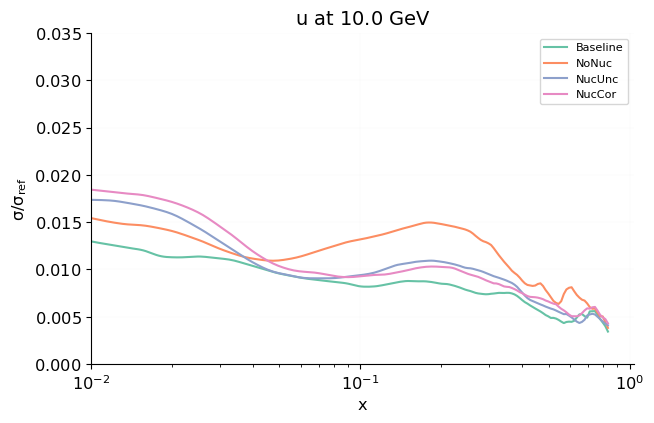}
\includegraphics[scale=0.47]{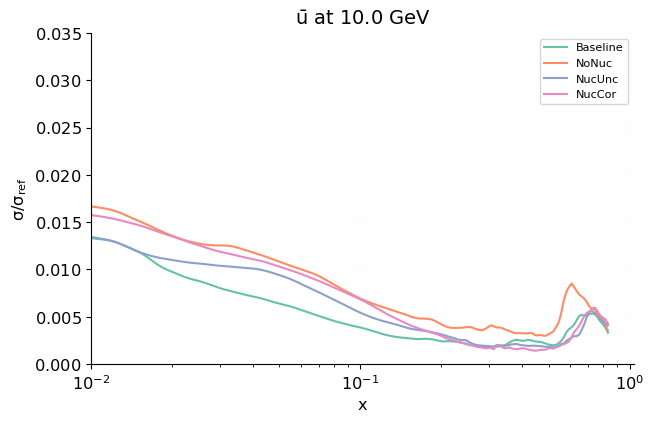}\\
\includegraphics[scale=0.47]{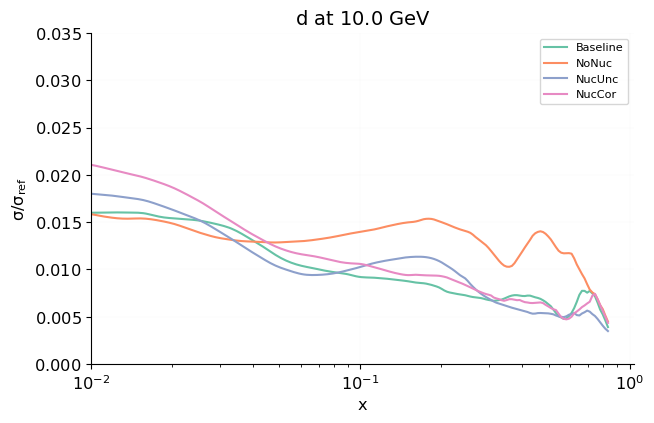}
\includegraphics[scale=0.47]{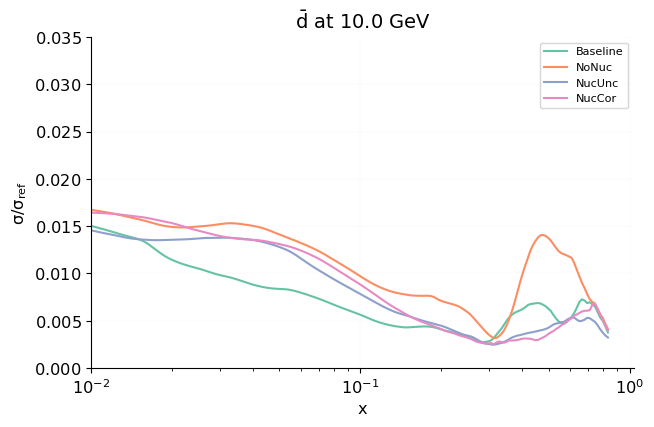}\\
\includegraphics[scale=0.47]{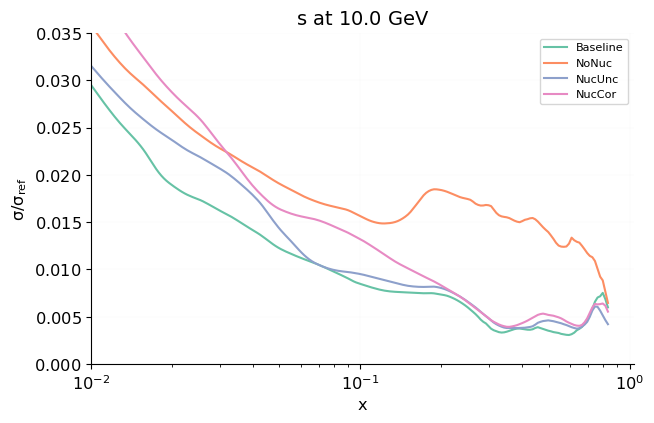}
\includegraphics[scale=0.47]{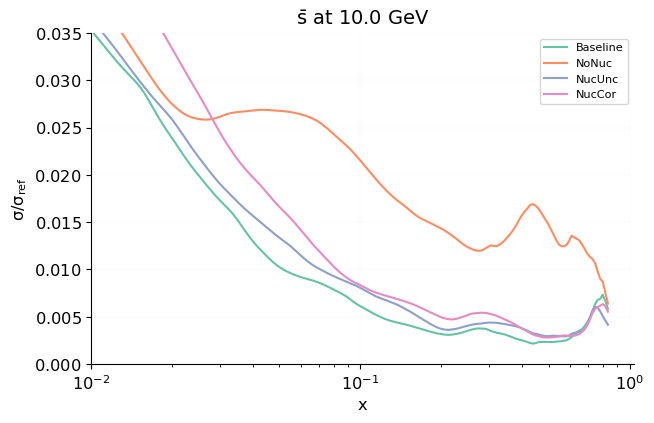}\\
\caption{Comparison of the absolute PDF uncertainty for the light quark and 
  antiquark flavours from the four fits performed in this analysis at 
  $Q=10$ GeV.}
\label{fig:xpdf_er}
\end{figure}

We now make the most important differences in the PDF central values and 
uncertainties among the four fits explicit.
In Fig.~\ref{fig:xpdf_cv_nonucl} we compare the light quark and antiquark PDFs
between the Baseline and the NoNuc fits.
In Fig.~\ref{fig:xpdf_cv_nucl} we compare the sea quark PDFs between the 
Baseline and either the NucUnc or the NucCor fits. 
In both figures, results are normalised to the Baseline fit.
In Fig.~\ref{fig:xpdf_er} we compare the absolute PDF uncertainty for the 
light quark and antiquark flavours from the four fits. 
All results are displayed at $Q=10$ GeV.

Inspecting Figs.~\ref{fig:xpdf_cv_nonucl}-\ref{fig:xpdf_er} we may draw a number of conclusions.
Firstly, the data taken on nuclear targets add a significant amount
of information to the global fit.
All light quark and antiquark PDFs are affected.
The effect on the $u$ and $d$ PDFs, which are expected to be already very
constrained by proton and deuteron data, consists of a slight distortion of
the corresponding central values, which however remain always 
included in the one-sigma uncertainty of the comparing fit.
More importantly, uncertainties are reduced by up to a factor of one third 
in the region $x\gtrsim 0.1$.
The effect on the sea quark PDFs is more pronounced.
Concerning central values, the nuclear data suppresses  
$\bar{u}$ and $\bar{d}$ PDFs below $x\sim 0.1$, and enhances them above 
$x\sim 0.1$. It also
suppresses $s$ and $\bar{s}$ above $x\sim 0.1$.
Uncertainties can be reduced down to two thirds (for $\bar{u}$ and $\bar{d}$) 
and to one quarter (for $s$ and $\bar{s}$) of the value obtained without
the nuclear data.
All these effects emphasise the constraining power of  
the nuclear data in a global fit, as already noted in Sect.~4.11 of 
Ref.~\cite{Ball:2017nwa}.

Second, the inclusion of theoretical uncertainties in the fit mostly affects
sea quark PDFs.
Central values are generally contained within the one-sigma 
uncertainty of the Baseline fit, that includes the nuclear data but 
does not include any theoretical
uncertainty, irrespective of the PDF flavour and of the prescription
used to estimate the theoretical covariance matrix.
The nature of the change in the central value is similar in both the NucUnc
and the NucCor fits.
For $\bar{u}$ and $\bar{d}$ PDFs, we observe an enhancement in the 
region $0.2\simeq x \simeq 0.3$ followed by a strong suppression for
$x\gtrsim 0.3$. 
For $s$ and $\bar{s}$ PDFs, we observe: a slight suppression at 
$x\lesssim 0.1-0.2$; an enhancement at $0.1-0.2\lesssim x \lesssim 0.4-0.5$;
and a strong suppression at $x\gtrsim 0.4-0.5$. 
Uncertainties are always increased when the theoretical covariance
matrix is included in the fit, irrespective of the way it is estimated, 
for all PDF flavours.
Such an increase is only marginally more apparent in the NucUnc fit, 
as a consequence of this being the more conservative estimate of the 
nuclear uncertainties.

All these effects lead us to conclude that theoretical uncertainties 
related to nuclear data are generally small in comparison to the 
experimental uncertainty of a typical global fit, in that deviations do not 
usually exceed one-sigma.
Nevertheless, slight distortions in the central values and increases
in the uncertainty bands (especially for $s$ and $\bar{s}$) become 
appreciable when theoretical uncertainties are taken into account.
The systematic inclusion of nuclear uncertainties in a global fit is thus 
advantageous whenever conservative predictions of sea quark PDFs are required.

\section{Impact on Phenomenology}
\label{sec:pheno}

As discussed in the previous section, the most sizeable impact of theoretical
uncertainties is on the light sea quark PDFs, which display slightly 
distorted central values, and appreciably inflated uncertainties in comparison
to the Baseline fit.
In this section, we study the implications of these effects on the 
sea quark asymmetry, and on the strangeness fraction of the proton,
including a possible asymmetry between $s$ and $\bar{s}$ PDFs.

\subsection{The Sea Quark Asymmetry}
\label{subsec:dbarubarratio}

A sizeable asymmetry in the antiup and antidown quark sea was observed long 
ago in DY first by the NA51~\cite{Baldit:1994jk} and then by the NuSea/E866 
experiments~\cite{Hawker:1998ty}.
Perturbatively, the number of antiup and antidown quarks in the proton 
is expected to be very nearly the same, because they originate primarily 
from the splitting of gluons into a quark-antiquark pair, and because 
their masses are very small in comparison to the confinement scale.
The observed asymmetry must therefore be explained by some
non-perturbative mechanism, which has been formulated in terms of 
various models over the years~\cite{Garvey:2001yq}.

\begin{figure}[!t]
\centering
\includegraphics[scale=0.225]{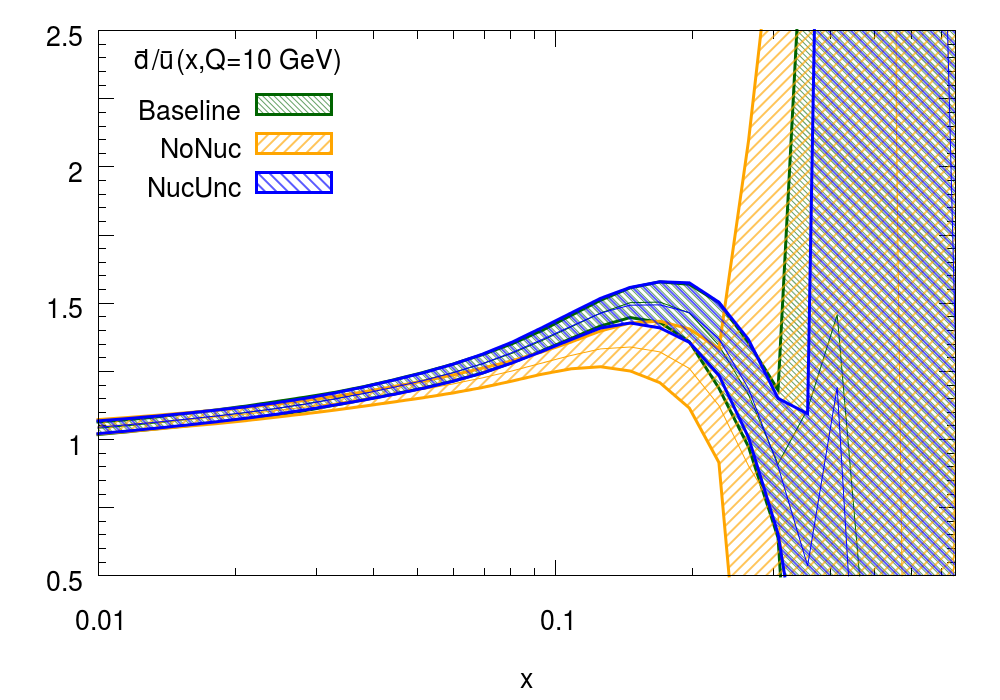}
\includegraphics[scale=0.225]{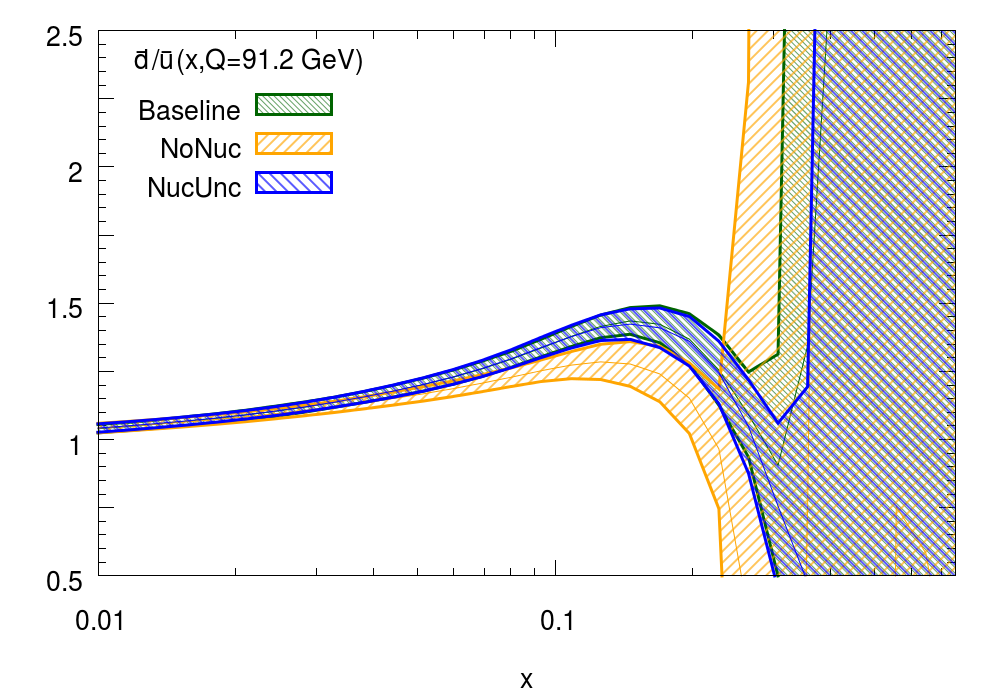}\\
\caption{The ratio $\bar{d}/\bar{u}$ as a function of $x$ at two 
representative values of $Q$ for the nuclear data ($Q=10$ GeV) and for the 
collider data ($Q=91.2$ GeV).}
\label{fig:du}
\end{figure}

While a complete understanding of this asymmetry is still lacking, we want
to analyse here the effect of the nuclear data and of their associated 
theoretical uncertainties due to nuclear effects on the PDF determination 
of the ratio $\bar{d}/\bar{u}$.
In Fig.~\ref{fig:du} we show the ratio $\bar{d}/\bar{u}$ as a function of $x$
at two representative values of $Q$ for the nuclear data ($Q=10$ GeV), and for 
the collider data ($Q=91.2$ GeV).
Results refer to the Baseline, NoNuc and NucUnc fits discussed
in Sect.~\ref{sec:results}.
We omit the NucCor result from Fig.~\ref{fig:du} for readability, as
it is almost indistinguishable from the NucUnc result.

Inspection of Fig.~\ref{fig:du} makes it apparent that the effect of nuclear
data on the $\bar{d}/\bar{u}$ ratio is significant, in particular in the region
$0.03\lesssim x \lesssim 0.3$.
In this region, the central value of the Baseline fit is enhanced by around 
two sigma with respect to the NoNuc fit.
The corresponding uncertainty bands do not show any significant difference in 
size, but they barely overlap.
The two fits differ by approximately $\sqrt{2}$ sigma.
The inclusion of the nuclear uncertainty, even at its most conservative, 
makes little difference to the $\bar{d}/\bar{u}$ ratio: the 
central value and the uncertainty of the NucUnc 
result are almost unchanged in comparison to the Baseline, and the ratio 
remains significantly larger than the NoNuc result.

\subsection{The Strange Content of the Proton Revisited}
\label{subsec:strange}

The size of the $s$ and $\bar{s}$ PDFs has recently been a source of 
some controversy.
Specifically, for many years the fraction of strange quarks in the proton 
\begin{equation}
R_s(x,Q^2) 
= 
\frac{s(x,Q^2) + \bar{s}(x,Q^2)}{\bar{d}(x,Q^2) + \bar{u}(x,Q^2)}\,,
\label{eq:Rs}
\end{equation}
and the corresponding momentum fraction
\begin{equation}
K_s(Q^2) 
= 
\frac{\int_0^1 dx \ x \left[ s(x,Q^2) + \bar{s}(x,Q^2) \right]}{
\int_0^1 dx \ x \left[ \bar{u}(x,Q^2) + \bar{d}(x,Q^2) \right]}
\label{eq:Ks}
\end{equation}
have been found to be smaller than one in all global PDF sets in which 
strange PDFs are fitted.
This result was mostly driven by NuTeV data, and has been understood to be 
due to the mass of the strange quark, which kinematically suppresses the 
production of $s$-$\bar{s}$ pairs.
This orthodoxy was challenged in Ref.~\cite{Aad:2012sb}, where, on the basis of 
ATLAS $W$ and $Z$ production data combined with HERA DIS data, it was 
claimed that the strange fraction $R_s$ is of order one.
The data of Ref.~\cite{Aad:2012sb} was later included in the NNPDF3.0
global analysis~\cite{Ball:2014uwa}, together with NuTeV data, to demonstrate 
that whereas the ATLAS data does favour a larger total strangeness, it has 
a moderate impact in the global fit due to its rather large uncertainties.
Furthermore, in a variant of the NNPDF3.0 analysis, if the $s$ and 
$\bar{s}$ PDFs were determined from HERA and ATLAS data only, the 
central value of $R_s$ is consistent with the conclusion of 
Ref.~\cite{Aad:2012sb}, and its uncertainty is large enough to make 
it compatible with the result of the global fit.
This state of affairs was reassessed in the NNDPF3.1 global
analysis~\cite{Ball:2017nwa}, where the ATLAS $W$ and $Z$ dataset was 
supplemented with the more precise measurements of Ref.~\cite{Aaboud:2016btc}.
They were found to enhance the total strangeness, consistently with that of 
Ref.~\cite{Aad:2012sb}, although their $\chi^2$ remained rather poor.
This suggested some residual tension in the preferred total strangeness 
between the ATLAS data and the rest of the dataset, specifically NuTeV.

\begin{figure}[!t]
\centering
\includegraphics[scale=0.225]{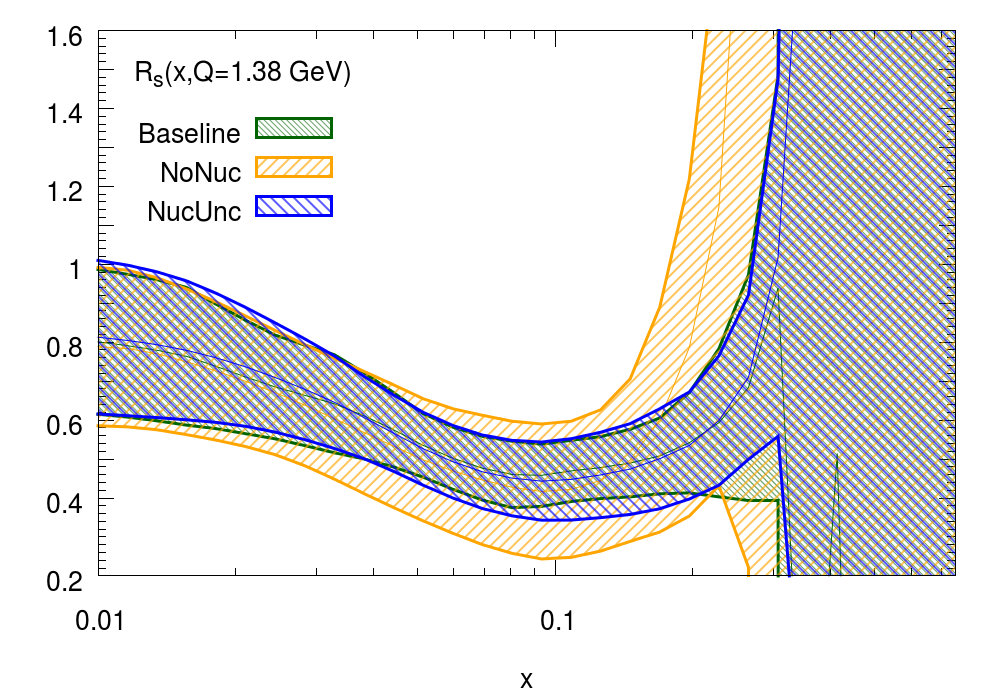}
\includegraphics[scale=0.225]{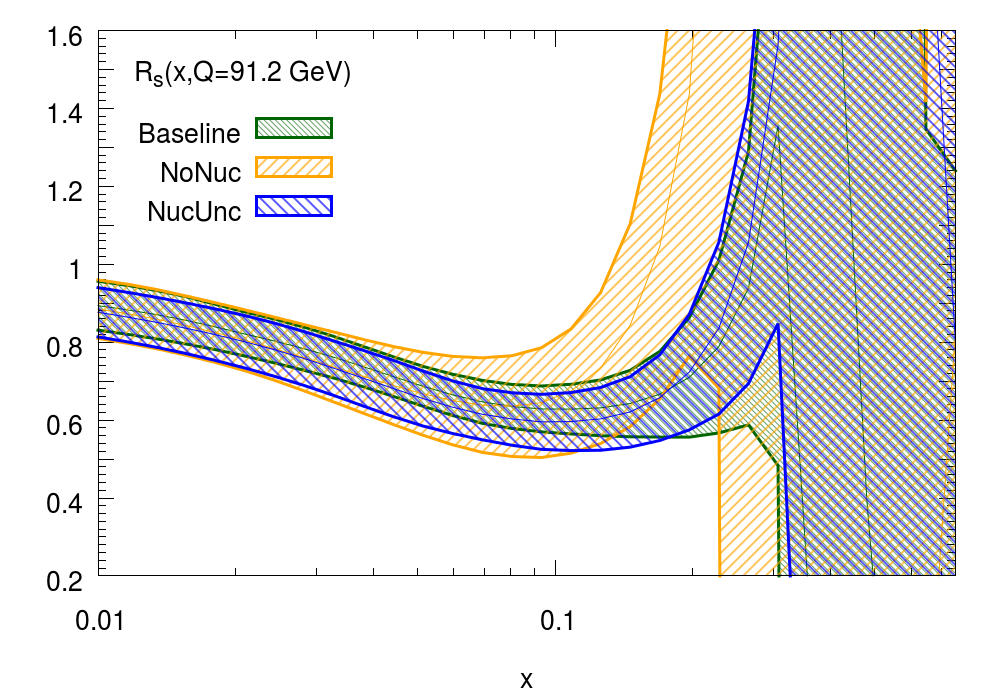}\\
\caption{The fraction $R_s$, Eq.~\eqref{eq:Rs}, as a function of $x$ at two 
representative values of $Q$ for the nuclear data ($Q=1.38$ GeV) and for the 
ATLAS $W$ $Z$ data ($Q=91.2$ GeV).}
\label{fig:Rs}
\end{figure}

We might wonder whether, since the NuTeV data were taken on an iron target, 
the inclusion of nuclear uncertainties might reconcile this discrepancy.
We therefore revisit the strange content of the proton, by computing the 
ratios $R_s$ and $K_s$, Eqs.~\eqref{eq:Rs}-\eqref{eq:Ks}, based on the four 
fits discused in the previous section at $Q=1.38$ GeV (below the charm 
threshold) and $Q=m_Z$, with $m_Z=91.2$ GeV the mass of the $Z$ boson. 
The values of $R_s$ at $x=0.023$ and of $K_s$ are collected in 
Table~\ref{tab:RsKs}, where the determination of Ref.~\cite{Aaboud:2016btc}
is also shown.
We display the ratio $R_s$ as a function of $x$ for the Baseline, NoNuc
and NucUnc fits in Fig.~\ref{fig:Rs}.
Again we omit the NucCor result from Fig.~\ref{fig:Rs} for readability, 
as it is almost indistinguishable from the NucUnc result.
The values of $Q$ and $x$ are the same as those used in the analysis  
of Ref.~\cite{Aaboud:2016btc}, where they were chosen to maximise sensitivity 
to either the nuclear (at low $Q$) or the collider (at high $Q$) data.

\begin{table}[!t]
\centering
\scriptsize
\renewcommand{\arraystretch}{1.13}
\begin{tabularx}{\textwidth}{lCCCCC}
\toprule
& Baseline
& NoNuc
& NucUnc
& NucCor
& Ref.~\cite{Aaboud:2016btc}\\
\midrule
  $R_s(x=0.023,Q=1.38\text{ GeV})$
& $+0.69    \pm 0.15$
& $+0.68    \pm 0.14$
& $+0.65    \pm 0.14$
& $+0.64    \pm 0.15$     
& $+1.13    \pm 0.11$\\
  $R_s(x=0.023,Q=91.2\text{ GeV})$
& $+0.81    \pm 0.05$
& $+0.79    \pm 0.07$
& $+0.79    \pm 0.06$
& $+0.77    \pm 0.06$
& ---      \\
\midrule
  $K_s(1.38\text{ GeV})$
& $+0.63     \pm 0.09$
& $+0.97     \pm 0.18$
& $+0.63     \pm 0.09$
& $+0.61     \pm 0.10$     
& --- \\
  $K_s(91.2\text{ GeV})$ 
& $+0.80     \pm 0.05$
& $+0.98     \pm 0.09$
& $+0.80     \pm 0.05$
& $+0.79     \pm 0.06$      
& --- \\      
\bottomrule
\end{tabularx}
\caption{The fractions $R_s$, Eq.~\eqref{eq:Rs}, and $K_s$, Eq.~\eqref{eq:Ks}, 
  for representative values of $x$ and $Q$.}
\label{tab:RsKs}
\end{table}

Inspection of Table~\ref{tab:RsKs} and Fig.~\ref{fig:Rs} makes it apparent that
the effect of nuclear data on $R_s$ at low values of $x$, namely at $x=0.023$, 
is negligible.
The central values and the uncertainties of $R_s$ are remarkably
stable across the four fits.
This is unsurprising, as $x=0.023$ is at the lower edge of the kinematic
region covered by the nuclear data, see Fig.~\ref{fig:dataset}.
At larger values of $x$, instead, particularly in the range 
$0.03\lesssim x\lesssim 0.2$, the nuclear data affects $R_s$ quite 
significantly, as is apparent from comparison of the NoNuc and Baseline fits.
In the Baseline (which includes the nuclear data), the uncertainty on 
$R_s$ is reduced by a factor of two without any apparent distortion of 
the central value.
Likewise, the effect of nuclear uncertainties is mostly apparent in a 
similar $x$ range.
If one compares the NucUnc and the Baseline fits, an increase of the 
uncertainty on $R_s$ by up to one third can be seen.
However, this effect remains moderate, and is mostly washed out when it is
integrated over the full range of $x$.
The value of $K_s$ is indeed almost unchanged by the inclusion of nuclear 
uncertainties in the Baseline fit, irrespective of whether they are 
implemented as the NucUnc or the NucCor fit, especially at high 
values of $Q$.

We therefore conclude that the inclusion of nuclear uncertainties does nothing 
to reconcile the residual tension between ATLAS and NuTeV data,
the reason being that they probe the strangeness in kinematic regions of
$x$ and $Q$ that barely overlap.
Further evidence of the limited interplay between ATLAS and NuTeV 
data is provided by the $\chi^2$ of the former, which remains poor 
for all the four fits considered in this analysis, see Table~\ref{tab:chi2}.
Achieving a better description of the ATLAS data or an improved determination 
of the strange content of the proton might require the 
inclusion of QCD corrections beyond NNLO and/or of electroweak corrections, or 
the analysis of other processes sensitive to $s$ and 
$\bar{s}$ PDFs, such as kaon production in semi-inclusive DIS.
All this remains beyond the scope of this work.

\begin{figure}[!t]
\centering
\includegraphics[scale=0.225]{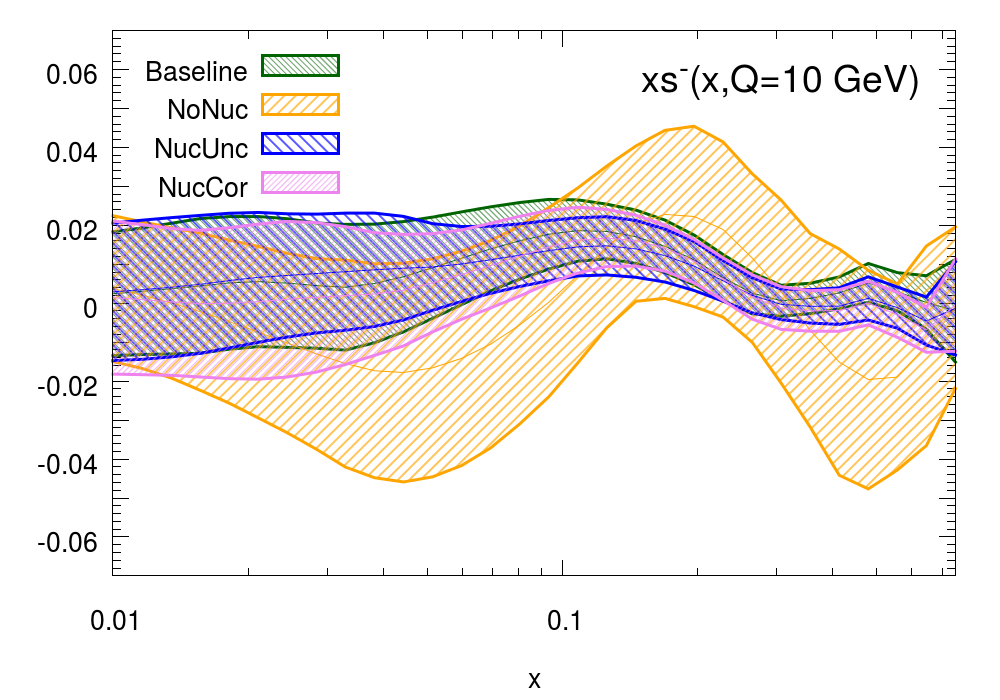}
\includegraphics[scale=0.225]{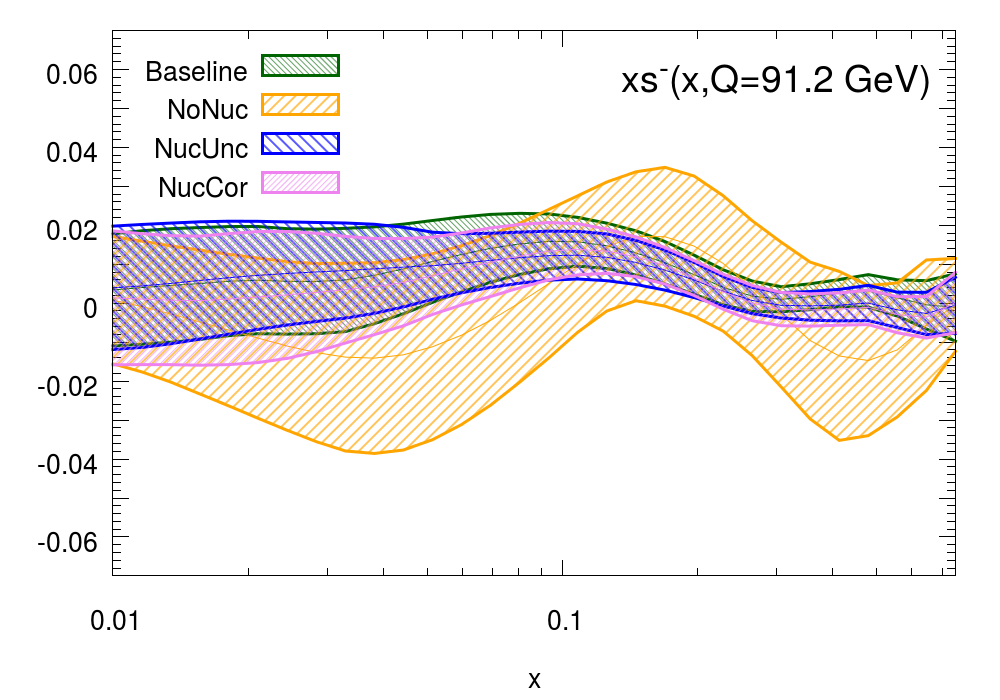}\\
\caption{The strange valence distribution $xs^-(x,Q)=x[s(x,Q)-\bar{s}(x,Q)]$
as a function of $x$ at two values of $Q$: a typical scale for the nuclear data 
($Q=10$ GeV), and for the collider data ($Q=91.2$ GeV).}
\label{fig:sval}
\end{figure}

Finally, we investigate the effect of the nuclear data on the strange valence
distribution $xs^-(x,Q)=x[s(x,Q)-\bar{s}(x,Q)]$, which we display as a function
of $x$ at two representative values of $Q$ in Fig.~\ref{fig:sval}.
Results are shown for each of the four fits performed in this analysis.
From Fig.~\ref{fig:sval}, we see once again the constraining power
of the nuclear data.
By comparing the Baseline and the NoNuc fits, it is apparent that the strange
valence distribution is almost unconstrained when the nuclear data is 
removed from the fit, with large uncertainty and completely unstable shape.

When the nuclear data is included, similar effects are observed whether nuclear 
uncertainties are implemented conservatively or as a correction.
Concerning the central values, in comparison to the Baseline fit the NucUnc 
and NucCor fits have a slightly suppressed valence distribution in the
region $x \lesssim 0.3$.
Overall, nuclear effects do not alter the asymmetry between 
$s$ and $\bar{s}$ PDFs.
Concerning the uncertainties themselves, both the NucUnc and the 
NucCor fits show an increased uncertainty in the strange valence 
distribution in the region $x \lesssim 0.3$, which is 
a little more pronounced for the the NucCor fit than the NucUnc fit, 
contrary to what would be expected if the small nuclear correction 
obtained from the nPDFs were a genuine effect.

In conclusion, nuclear effects have negligible impact on the 
$\bar{d}-\bar{u}$ asymmetry, on the total strangeness, and on the 
$s -\bar{s}$ asymmetry. 
We do not observe any evidence in support of the use of nuclear corrections
(NucCor): in the global proton fit, the fit quality to the nuclear 
datasets (and the overall fit quality) is always a little worse when 
nuclear corrections are implemented. This may be due 
to the slight inconsistencies between the different sets of nPDFs, visible in 
Fig.~\ref{fig:nPDFs}.
The use of the more conservative nuclear uncertainties (NucUnc) in global 
proton PDF fits is thus the recommended option, 
at least until more reliable nPDFs become available.

\section{Summary and Outlook}
\label{sec:conclusions}

In this paper we revisited the r\^ole of the nuclear data 
commonly used in a global determination of proton PDFs.
Specifically, we considered: DIS data taken with Pb and Fe targets, from 
CHORUS and NuTeV experiments, respectively; and DY data taken with a Cu 
target, from the E605 experiment.
We studied the fit quality and the stability of the proton PDFs obtained in 
the framework of the NNDPF3.1 global analysis, by comparing a series of 
determinations: one in which the nuclear dataset is removed from the fit; 
one in which it is included without any nuclear correction (the baseline fit); 
and two in which it is included with a theoretical uncertainty that takes 
into account nuclear effects.

The two determinations which included a theoretical uncertainty were realised 
by constructing a theoretical covariance matrix that was added to the 
experimental covariance matrix, both when generating data replicas 
and when fitting PDFs. 
These covariance matrices were constructed using a Monte Carlo
ensemble of nuclear PDFs, determined from a wide set of measurements 
using nuclear data.
In the first determination, the theoretical covariance matrix elements
were constructed by finding the difference between each nuclear PDF
replica and a \textit{central proton} PDF, and then taking an average 
over replicas.
This gives a conservative estimate of the nuclear uncertainty, increasing 
uncertainties overall, and deweighting the nuclear data in the fit.
In the second determination, the theoretical covariance matrix elements
were constructed by finding the difference between each nuclear PDF
replica and the \textit{central nuclear} PDF.
Additionally, the theoretical prediction was shifted by the difference 
between the predictions made with central nuclear and proton PDFs. This 
correction procedure takes the nuclear effects and their uncertainties 
as determined by the nuclear fits at face value, but is less conservative 
than the first prescription.

We confirm that the nuclear dataset contains substantial information on the 
proton PDFs, even when the uncertainties due to nuclear effects 
are taken into account. In particular it provides an important constraint 
on the light sea quark PDFs, consistent with constraints from LHC data, as we 
explicitly demonstrated by inspecting the individual PDFs, the 
$\bar{d}/\bar{u}$ ratio, the strangeness fractions $R_s$ and $K_s$, 
and the strange valence distribution $s-\bar{s}$.
Therefore, it should not be dropped from current global PDF fits.

A conservative estimate of the additional theoretical uncertainty due to 
the use of a nuclear rather than a proton target in these measurements 
gives only small changes in the central values of the proton PDFs, with  
a slight increase in their overall uncertainties.
In particular, nuclear effects are insufficient to explain any residual 
tension in the global fit between fixed target and ATLAS determinations 
of the strangeness content of the proton. 
This is largely because the corresponding measurements are sensitive to 
different kinematic regions that have a limited interplay.
We nevertheless recommend that nuclear uncertainties are always included 
in future global fits, to eliminate any slight bias, and as a precaution 
against underestimation of uncertainties.

We should emphasise that all our results were determined from 
the most recent publically available nuclear PDFs.
Despite the fact that these are obtained from a global analysis of 
experimental data taken  
in a wide variety of processes, including DIS, DY and $pN$ collisions, 
some inconsistencies between the nPDF sets, in particular in the 
estimation of the uncertainties, were observed.
This suggests that nPDFs are not yet sufficiently reliable to justify
the use of a nuclear correction in the fit. 
Indeed, when we attempted to use them to give a nuclear correction to 
the predictions with proton PDFs, the fit quality was always a 
little worse than that obtained without nuclear corrections. 
Even so, the effect on the proton PDFs, in particular the $\bar{d}/\bar{u}$ 
ratio, the total strangeness $s+\bar{s}$, and the valence strange distribution 
$s-\bar{s}$ was negligible.
Nevertheless, we recommend that when estimating nuclear effects using 
current nPDF sets, the more conservative approach is adopted, in which 
nuclear effects give an additional uncertainty, but not a correction. 

Our work can be extended in various different directions.
First, we intend to reconsider our results when newer more reliable 
nPDF sets become available.
In this respect, a consistent determination based on the NNPDF 
methodology~\cite{Khalek:2018bbv} would be particularly helpful, both because 
it would give a more reliable assessment of the uncertainties on nPDFs, and 
because a  consistent treatment, determining proton PDFs, nPDFs and 
nuclear corrections iteratively, would then be possible.

Second, our analysis can be extended to deuterium data.
In principle, this suffers from similar 
uncertainties as the nuclear data considered here, though they are 
expected to be much smaller.
Theoretical uncertainties might be estimated by studying the spread between
various model predictions for nuclear effects, or by attempting to 
determine them empirically from the data, in iterating them to consistency 
with the proton data.

Finally, the general method of accounting for theoretical uncertainties
in a PDF fit by estimating a theoretical covariance matrix that is added to the 
experimental one in the definition of the $\chi^2$ can have many other
applications~\cite{Ball:2018odr}: missing higher-order 
uncertainties~\cite{Pearson:2018tim}, higher-twist uncertainties, and 
fragmentation function uncertainties in the analysis
of semi-inclusive data.
In this last respect, the analysis of kaon production data might be useful 
to obtain more information on the strangeness content of the proton.

\vspace{0.5cm}

The PDF sets presented in this work are available in the {\tt LHAPDF} 
format~\cite{Buckley:2014ana} from the authors upon request.

\section*{Acknowledgements}

We thank R.~Sassot for providing us with the nuclear PDF sets of 
Ref.~\cite{deFlorian:2011fp}.
We acknowledge useful discussions on the inclusion of theoretical uncertainties
in PDF fits with our colleagues in the NNPDF Collaboration.
The authors are supported by the UK Science and Technology Facility 
Council through grant ST/P000630/1. 
E.R.N. is also supported by the European Commission through the 
Marie Sk\l odowska-Curie Action ParDHonS\_FFs.TMDs (grant number 752748).

\bibliography{nucl-corrs}

\providecommand{\href}[2]{#2}\begingroup\raggedright\begin{thebibliography}{100}

\bibitem{Butterworth:2015oua}
J.~Butterworth et~al., {\it {PDF4LHC recommendations for LHC Run II}},  {\em J.
  Phys.} {\bf G43} (2016) 023001, [\href{http://arxiv.org/abs/1510.03865}{{\tt
  arXiv:1510.03865}}].

\bibitem{Gao:2017yyd}
J.~Gao, L.~Harland-Lang, and J.~Rojo, {\it {The Structure of the Proton in the
  LHC Precision Era}},  \href{http://arxiv.org/abs/1709.04922}{{\tt
  arXiv:1709.04922}}.

\bibitem{Ball:2017nwa}
{\bf NNPDF} Collaboration, R.~D. Ball et~al., {\it {Parton distributions from
  high-precision collider data}},  {\em Eur. Phys. J.} {\bf C77} (2017), no.~10
  663, [\href{http://arxiv.org/abs/1706.00428}{{\tt arXiv:1706.00428}}].

\bibitem{Harland-Lang:2014zoa}
L.~A. Harland-Lang, A.~D. Martin, P.~Motylinski, and R.~S. Thorne, {\it {Parton
  distributions in the LHC era: MMHT 2014 PDFs}},  {\em Eur. Phys. J.} {\bf
  C75} (2015), no.~5 204, [\href{http://arxiv.org/abs/1412.3989}{{\tt
  arXiv:1412.3989}}].

\bibitem{Dulat:2015mca}
S.~Dulat, T.-J. Hou, J.~Gao, M.~Guzzi, J.~Huston, P.~Nadolsky, J.~Pumplin,
  C.~Schmidt, D.~Stump, and C.~P. Yuan, {\it {New parton distribution functions
  from a global analysis of quantum chromodynamics}},  {\em Phys. Rev.} {\bf
  D93} (2016), no.~3 033006, [\href{http://arxiv.org/abs/1506.07443}{{\tt
  arXiv:1506.07443}}].

\bibitem{Alekhin:2017kpj}
S.~Alekhin, J.~Blümlein, S.~Moch, and R.~Placakyte, {\it {Parton distribution
  functions, $\alpha_s$, and heavy-quark masses for LHC Run II}},  {\em Phys.
  Rev.} {\bf D96} (2017), no.~1 014011,
  [\href{http://arxiv.org/abs/1701.05838}{{\tt arXiv:1701.05838}}].

\bibitem{Accardi:2016qay}
A.~Accardi, L.~T. Brady, W.~Melnitchouk, J.~F. Owens, and N.~Sato, {\it
  {Constraints on large-$x$ parton distributions from new weak boson production
  and deep-inelastic scattering data}},  {\em Phys. Rev.} {\bf D93} (2016),
  no.~11 114017, [\href{http://arxiv.org/abs/1602.03154}{{\tt
  arXiv:1602.03154}}].

\bibitem{Ball:2009mk}
{\bf NNPDF} Collaboration, R.~D. Ball, L.~Del~Debbio, S.~Forte, A.~Guffanti,
  J.~I. Latorre, A.~Piccione, J.~Rojo, and M.~Ubiali, {\it {Precision
  determination of electroweak parameters and the strange content of the proton
  from neutrino deep-inelastic scattering}},  {\em Nucl. Phys.} {\bf B823}
  (2009) 195--233, [\href{http://arxiv.org/abs/0906.1958}{{\tt
  arXiv:0906.1958}}].

\bibitem{Ball:2013gsa}
{\bf NNPDF} Collaboration, R.~D. Ball, V.~Bertone, L.~Del~Debbio, S.~Forte,
  A.~Guffanti, J.~Rojo, and M.~Ubiali, {\it {Theoretical issues in PDF
  determination and associated uncertainties}},  {\em Phys. Lett.} {\bf B723}
  (2013) 330--339, [\href{http://arxiv.org/abs/1303.1189}{{\tt
  arXiv:1303.1189}}].

\bibitem{Owens:2012bv}
J.~F. Owens, A.~Accardi, and W.~Melnitchouk, {\it {Global parton distributions
  with nuclear and finite-$Q^2$ corrections}},  {\em Phys. Rev.} {\bf D87}
  (2013), no.~9 094012, [\href{http://arxiv.org/abs/1212.1702}{{\tt
  arXiv:1212.1702}}].

\bibitem{Ball:2018odr}
R.~D. Ball and A.~Deshpande, {\it {The Proton Spin, Semi-Inclusive processes,
  and a future Electron Ion Collider}},
\newblock 2018.
\newblock \href{http://arxiv.org/abs/1801.04842}{{\tt arXiv:1801.04842}}.

\bibitem{Ball:2008by}
{\bf NNPDF} Collaboration, R.~D. Ball, L.~Del~Debbio, S.~Forte, A.~Guffanti,
  J.~I. Latorre, A.~Piccione, J.~Rojo, and M.~Ubiali, {\it {A Determination of
  parton distributions with faithful uncertainty estimation}},  {\em Nucl.
  Phys.} {\bf B809} (2009) 1--63, [\href{http://arxiv.org/abs/0808.1231}{{\tt
  arXiv:0808.1231}}]. [Erratum: Nucl. Phys.B816,293(2009)].

\bibitem{DAgostini:2003syq}
G.~D'Agostini, {\em {Bayesian reasoning in data analysis: A critical
  introduction}}.
\newblock 2003.

\bibitem{Ball:2009qv}
{\bf NNPDF} Collaboration, R.~D. Ball, L.~Del~Debbio, S.~Forte, A.~Guffanti,
  J.~I. Latorre, J.~Rojo, and M.~Ubiali, {\it {Fitting Parton Distribution Data
  with Multiplicative Normalization Uncertainties}},  {\em JHEP} {\bf 05}
  (2010) 075, [\href{http://arxiv.org/abs/0912.2276}{{\tt arXiv:0912.2276}}].

\bibitem{Onengut:2005kv}
{\bf CHORUS} Collaboration, G.~Onengut et~al., {\it {Measurement of nucleon
  structure functions in neutrino scattering}},  {\em Phys. Lett.} {\bf B632}
  (2006) 65--75.

\bibitem{Goncharov:2001qe}
{\bf NuTeV} Collaboration, M.~Goncharov et~al., {\it {Precise measurement of
  dimuon production cross-sections in muon neutrino Fe and muon anti-neutrino
  Fe deep inelastic scattering at the Tevatron}},  {\em Phys. Rev.} {\bf D64}
  (2001) 112006, [\href{http://arxiv.org/abs/hep-ex/0102049}{{\tt
  hep-ex/0102049}}].

\bibitem{Mason:2006qa}
D.~A. Mason, {\em {Measurement of the strange - antistrange asymmetry at NLO in
  QCD from NuTeV dimuon data}}.
\newblock PhD thesis, Oregon U., 2006.

\bibitem{Moreno:1990sf}
G.~Moreno et~al., {\it {Dimuon production in proton - copper collisions at
  $\sqrt{s}$ = 38.8-GeV}},  {\em Phys. Rev.} {\bf D43} (1991) 2815--2836.

\bibitem{Berge:1989hr}
J.~P. Berge et~al., {\it {A Measurement of Differential Cross-Sections and
  Nucleon Structure Functions in Charged Current Neutrino Interactions on
  Iron}},  {\em Z. Phys.} {\bf C49} (1991) 187--224.

\bibitem{Guffanti:2010yu}
A.~Guffanti and J.~Rojo, {\it {Top production at the LHC: The Impact of PDF
  uncertainties and correlations}},  {\em Nuovo Cim.} {\bf C033} (2010), no.~4
  65--72, [\href{http://arxiv.org/abs/1008.4671}{{\tt arXiv:1008.4671}}].

\bibitem{Arneodo:1992wf}
M.~Arneodo, {\it {Nuclear effects in structure functions}},  {\em Phys. Rept.}
  {\bf 240} (1994) 301--393.

\bibitem{deFlorian:2011fp}
D.~de~Florian, R.~Sassot, P.~Zurita, and M.~Stratmann, {\it {Global Analysis of
  Nuclear Parton Distributions}},  {\em Phys. Rev.} {\bf D85} (2012) 074028,
  [\href{http://arxiv.org/abs/1112.6324}{{\tt arXiv:1112.6324}}].

\bibitem{Kovarik:2015cma}
K.~Kovarik et~al., {\it {nCTEQ15 - Global analysis of nuclear parton
  distributions with uncertainties in the CTEQ framework}},  {\em Phys. Rev.}
  {\bf D93} (2016), no.~8 085037, [\href{http://arxiv.org/abs/1509.00792}{{\tt
  arXiv:1509.00792}}].

\bibitem{Eskola:2016oht}
K.~J. Eskola, P.~Paakkinen, H.~Paukkunen, and C.~A. Salgado, {\it {EPPS16:
  Nuclear parton distributions with LHC data}},  {\em Eur. Phys. J.} {\bf C77}
  (2017), no.~3 163, [\href{http://arxiv.org/abs/1612.05741}{{\tt
  arXiv:1612.05741}}].

\bibitem{Paukkunen:2017bbm}
H.~Paukkunen, {\it {Status of nuclear PDFs after the first LHC p–Pb run}},
  {\em Nucl. Phys.} {\bf A967} (2017) 241--248,
  [\href{http://arxiv.org/abs/1704.04036}{{\tt arXiv:1704.04036}}].

\bibitem{Paukkunen:2018kmm}
H.~Paukkunen, {\it {Nuclear PDFs Today}},  in {\em {9th International
  Conference on Hard and Electromagnetic Probes of High-Energy Nuclear
  Collisions: Hard Probes 2018 (HP2018) Aix-Les-Bains, Savoie, France, October
  1-5, 2018}}, 2018.
\newblock \href{http://arxiv.org/abs/1811.01976}{{\tt arXiv:1811.01976}}.

\bibitem{Arneodo:1996qe}
{\bf New Muon} Collaboration, M.~Arneodo et~al., {\it {Measurement of the
  proton and deuteron structure functions, F2(p) and F2(d), and of the ratio
  sigma-L / sigma-T}},  {\em Nucl. Phys.} {\bf B483} (1997) 3--43,
  [\href{http://arxiv.org/abs/hep-ph/9610231}{{\tt hep-ph/9610231}}].

\bibitem{Amaudruz:1995tq}
{\bf New Muon} Collaboration, P.~Amaudruz et~al., {\it {A Reevaluation of the
  nuclear structure function ratios for D, He, Li-6, C and Ca}},  {\em Nucl.
  Phys.} {\bf B441} (1995) 3--11,
  [\href{http://arxiv.org/abs/hep-ph/9503291}{{\tt hep-ph/9503291}}].

\bibitem{Gomez:1993ri}
J.~Gomez et~al., {\it {Measurement of the A-dependence of deep inelastic
  electron scattering}},  {\em Phys. Rev.} {\bf D49} (1994) 4348--4372.

\bibitem{Airapetian:2002fx}
{\bf HERMES} Collaboration, A.~Airapetian et~al., {\it {Measurement of R =
  sigma(L) / sigma(T) in deep inelastic scattering on nuclei}},
  \href{http://arxiv.org/abs/hep-ex/0210068}{{\tt hep-ex/0210068}}.

\bibitem{Arneodo:1995cs}
{\bf New Muon} Collaboration, M.~Arneodo et~al., {\it {The Structure Function
  ratios F2(li) / F2(D) and F2(C) / F2(D) at small x}},  {\em Nucl. Phys.} {\bf
  B441} (1995) 12--30, [\href{http://arxiv.org/abs/hep-ex/9504002}{{\tt
  hep-ex/9504002}}].

\bibitem{Adams:1995is}
{\bf E665} Collaboration, M.~R. Adams et~al., {\it {Shadowing in inelastic
  scattering of muons on carbon, calcium and lead at low x(Bj)}},  {\em Z.
  Phys.} {\bf C67} (1995) 403--410,
  [\href{http://arxiv.org/abs/hep-ex/9505006}{{\tt hep-ex/9505006}}].

\bibitem{Ashman:1988bf}
{\bf European Muon} Collaboration, J.~Ashman et~al., {\it {Measurement of the
  Ratios of Deep Inelastic Muon-Nucleus Cross-Sections on Various Nuclei
  Compared to Deuterium}},  {\em Phys. Lett.} {\bf B202} (1988) 603--610.

\bibitem{Bari:1985ga}
{\bf BCDMS} Collaboration, G.~Bari et~al., {\it {A Measurement of Nuclear
  Effects in Deep Inelastic Muon Scattering on Deuterium, Nitrogen and Iron
  Targets}},  {\em Phys. Lett.} {\bf 163B} (1985) 282.

\bibitem{Bodek:1983qn}
A.~Bodek et~al., {\it {Electron Scattering from Nuclear Targets and Quark
  Distributions in Nuclei}},  {\em Phys. Rev. Lett.} {\bf 50} (1983) 1431.

\bibitem{Benvenuti:1987az}
{\bf BCDMS} Collaboration, A.~C. Benvenuti et~al., {\it {Nuclear Effects in
  Deep Inelastic Muon Scattering on Deuterium and Iron Targets}},  {\em Phys.
  Lett.} {\bf B189} (1987) 483--487.

\bibitem{Ashman:1992kv}
{\bf European Muon} Collaboration, J.~Ashman et~al., {\it {A Measurement of the
  ratio of the nucleon structure function in copper and deuterium}},  {\em Z.
  Phys.} {\bf C57} (1993) 211--218.

\bibitem{Arneodo:1996rv}
{\bf New Muon} Collaboration, M.~Arneodo et~al., {\it {The A dependence of the
  nuclear structure function ratios}},  {\em Nucl. Phys.} {\bf B481} (1996)
  3--22.

\bibitem{Arneodo:1996ru}
{\bf New Muon} Collaboration, M.~Arneodo et~al., {\it {The Q**2 dependence of
  the structure function ratio F2 Sn / F2 C and the difference R Sn - R C in
  deep inelastic muon scattering}},  {\em Nucl. Phys.} {\bf B481} (1996)
  23--39.

\bibitem{Tzanov:2005kr}
{\bf NuTeV} Collaboration, M.~Tzanov et~al., {\it {Precise measurement of
  neutrino and anti-neutrino differential cross sections}},  {\em Phys. Rev.}
  {\bf D74} (2006) 012008, [\href{http://arxiv.org/abs/hep-ex/0509010}{{\tt
  hep-ex/0509010}}].

\bibitem{Alde:1990im}
D.~M. Alde et~al., {\it {Nuclear dependence of dimuon production at 800-GeV.
  FNAL-772 experiment}},  {\em Phys. Rev. Lett.} {\bf 64} (1990) 2479--2482.

\bibitem{Vasilev:1999fa}
{\bf NuSea} Collaboration, M.~A. Vasilev et~al., {\it {Parton energy loss
  limits and shadowing in Drell-Yan dimuon production}},  {\em Phys. Rev.
  Lett.} {\bf 83} (1999) 2304--2307,
  [\href{http://arxiv.org/abs/hep-ex/9906010}{{\tt hep-ex/9906010}}].

\bibitem{Adler:2006wg}
{\bf PHENIX} Collaboration, S.~S. Adler et~al., {\it {Centrality dependence of
  pi0 and eta production at large transverse momentum in s(NN)**(1/2) = 200-GeV
  d+Au collisions}},  {\em Phys. Rev. Lett.} {\bf 98} (2007) 172302,
  [\href{http://arxiv.org/abs/nucl-ex/0610036}{{\tt nucl-ex/0610036}}].

\bibitem{Abelev:2009hx}
{\bf STAR} Collaboration, B.~I. Abelev et~al., {\it {Inclusive $\pi^0$, $\eta$,
  and direct photon production at high transverse momentum in $p+p$ and $d+$Au
  collisions at $\sqrt{s_{NN}}=200$ GeV}},  {\em Phys. Rev.} {\bf C81} (2010)
  064904, [\href{http://arxiv.org/abs/0912.3838}{{\tt arXiv:0912.3838}}].

\bibitem{Bordalo:1987cs}
{\bf NA10} Collaboration, P.~Bordalo et~al., {\it {Nuclear Effects on the
  Nucleon Structure Functions in Hadronic High Mass Dimuon Production}},  {\em
  Phys. Lett.} {\bf B193} (1987) 368.

\bibitem{Heinrich:1989cp}
J.~G. Heinrich et~al., {\it {Measurement of the Ratio of Sea to Valence Quarks
  in the Nucleon}},  {\em Phys. Rev. Lett.} {\bf 63} (1989) 356--359.

\bibitem{Badier:1981ci}
{\bf NA3} Collaboration, J.~Badier et~al., {\it {Test of Nuclear Effects in
  Hadronic Dimuon Production}},  {\em Phys. Lett.} {\bf 104B} (1981) 335.
  [,807(1981)].

\bibitem{Khachatryan:2015hha}
{\bf CMS} Collaboration, V.~Khachatryan et~al., {\it {Study of W boson
  production in pPb collisions at $\sqrt{s_{\mathrm{NN}}} =$ 5.02 TeV}},  {\em
  Phys. Lett.} {\bf B750} (2015) 565--586,
  [\href{http://arxiv.org/abs/1503.05825}{{\tt arXiv:1503.05825}}].

\bibitem{Khachatryan:2015pzs}
{\bf CMS} Collaboration, V.~Khachatryan et~al., {\it {Study of Z boson
  production in pPb collisions at $\sqrt {s_{NN}} = 5.02$ TeV}},  {\em Phys.
  Lett.} {\bf B759} (2016) 36--57, [\href{http://arxiv.org/abs/1512.06461}{{\tt
  arXiv:1512.06461}}].

\bibitem{Aad:2015gta}
{\bf ATLAS} Collaboration, G.~Aad et~al., {\it {$Z$ boson production in $p+$Pb
  collisions at $\sqrt{s_{NN}}=5.02$ TeV measured with the ATLAS detector}},
  {\em Phys. Rev.} {\bf C92} (2015), no.~4 044915,
  [\href{http://arxiv.org/abs/1507.06232}{{\tt arXiv:1507.06232}}].

\bibitem{Sassot:2009sh}
R.~Sassot, M.~Stratmann, and P.~Zurita, {\it {Fragmentations Functions in
  Nuclear Media}},  {\em Phys. Rev.} {\bf D81} (2010) 054001,
  [\href{http://arxiv.org/abs/0912.1311}{{\tt arXiv:0912.1311}}].

\bibitem{Sassot:PC}
R.~Sassot.
\newblock Private communication.

\bibitem{Watt:2012tq}
G.~Watt and R.~S. Thorne, {\it {Study of Monte Carlo approach to experimental
  uncertainty propagation with MSTW 2008 PDFs}},  {\em JHEP} {\bf 08} (2012)
  052, [\href{http://arxiv.org/abs/1205.4024}{{\tt arXiv:1205.4024}}].

\bibitem{Hou:2016sho}
T.-J. Hou et~al., {\it {Reconstruction of Monte Carlo replicas from Hessian
  parton distributions}},  {\em JHEP} {\bf 03} (2017) 099,
  [\href{http://arxiv.org/abs/1607.06066}{{\tt arXiv:1607.06066}}].

\bibitem{Carrazza:2015hva}
S.~Carrazza, J.~I. Latorre, J.~Rojo, and G.~Watt, {\it {A compression algorithm
  for the combination of PDF sets}},  {\em Eur. Phys. J.} {\bf C75} (2015) 474,
  [\href{http://arxiv.org/abs/1504.06469}{{\tt arXiv:1504.06469}}].

\bibitem{Owens:2007kp}
J.~F. Owens, J.~Huston, C.~E. Keppel, S.~Kuhlmann, J.~G. Morfin, F.~Olness,
  J.~Pumplin, and D.~Stump, {\it {The Impact of new neutrino DIS and Drell-Yan
  data on large-x parton distributions}},  {\em Phys. Rev.} {\bf D75} (2007)
  054030, [\href{http://arxiv.org/abs/hep-ph/0702159}{{\tt hep-ph/0702159}}].

\bibitem{Martin:2009iq}
A.~D. Martin, W.~J. Stirling, R.~S. Thorne, and G.~Watt, {\it {Parton
  distributions for the LHC}},  {\em Eur. Phys. J.} {\bf C63} (2009) 189--285,
  [\href{http://arxiv.org/abs/0901.0002}{{\tt arXiv:0901.0002}}].

\bibitem{Bertone:2013vaa}
V.~Bertone, S.~Carrazza, and J.~Rojo, {\it {APFEL: A PDF Evolution Library with
  QED corrections}},  {\em Comput. Phys. Commun.} {\bf 185} (2014) 1647--1668,
  [\href{http://arxiv.org/abs/1310.1394}{{\tt arXiv:1310.1394}}].

\bibitem{zahari_kassabov_2019_2571601}
Z.~Kassabov, {\it {Reportengine: A framework for declarative data analysis}},
  Feb., 2019.

\bibitem{Arneodo:1996kd}
{\bf New Muon} Collaboration, M.~Arneodo et~al., {\it {Accurate measurement of
  F2(d) / F2(p) and R**d - R**p}},  {\em Nucl. Phys.} {\bf B487} (1997) 3--26,
  [\href{http://arxiv.org/abs/hep-ex/9611022}{{\tt hep-ex/9611022}}].

\bibitem{Benvenuti:1989rh}
{\bf BCDMS} Collaboration, A.~C. Benvenuti et~al., {\it {A High Statistics
  Measurement of the Proton Structure Functions F(2) (x, Q**2) and R from Deep
  Inelastic Muon Scattering at High Q**2}},  {\em Phys. Lett.} {\bf B223}
  (1989) 485--489.

\bibitem{Benvenuti:1989fm}
{\bf BCDMS} Collaboration, A.~C. Benvenuti et~al., {\it {A High Statistics
  Measurement of the Deuteron Structure Functions F2 (X, $Q^2$) and R From Deep
  Inelastic Muon Scattering at High $Q^2$}},  {\em Phys. Lett.} {\bf B237}
  (1990) 592--598.

\bibitem{Whitlow:1991uw}
L.~W. Whitlow, E.~M. Riordan, S.~Dasu, S.~Rock, and A.~Bodek, {\it {Precise
  measurements of the proton and deuteron structure functions from a global
  analysis of the SLAC deep inelastic electron scattering cross-sections}},
  {\em Phys. Lett.} {\bf B282} (1992) 475--482.

\bibitem{Abramowicz:2015mha}
{\bf ZEUS, H1} Collaboration, H.~Abramowicz et~al., {\it {Combination of
  measurements of inclusive deep inelastic ${e^{\pm }p}$ scattering cross
  sections and QCD analysis of HERA data}},  {\em Eur. Phys. J.} {\bf C75}
  (2015), no.~12 580, [\href{http://arxiv.org/abs/1506.06042}{{\tt
  arXiv:1506.06042}}].

\bibitem{Abramowicz:1900rp}
{\bf ZEUS, H1} Collaboration, H.~Abramowicz et~al., {\it {Combination and QCD
  Analysis of Charm Production Cross Section Measurements in Deep-Inelastic ep
  Scattering at HERA}},  {\em Eur. Phys. J.} {\bf C73} (2013), no.~2 2311,
  [\href{http://arxiv.org/abs/1211.1182}{{\tt arXiv:1211.1182}}].

\bibitem{Webb:2003ps}
{\bf NuSea} Collaboration, J.~C. Webb et~al., {\it {Absolute Drell-Yan dimuon
  cross-sections in 800 GeV / c pp and pd collisions}},
  \href{http://arxiv.org/abs/hep-ex/0302019}{{\tt hep-ex/0302019}}.

\bibitem{Webb:2003bj}
J.~C. Webb, {\em {Measurement of continuum dimuon production in 800-GeV/C
  proton nucleon collisions}}.
\newblock PhD thesis, New Mexico State U., 2003.
\newblock \href{http://arxiv.org/abs/hep-ex/0301031}{{\tt hep-ex/0301031}}.

\bibitem{Towell:2001nh}
{\bf NuSea} Collaboration, R.~S. Towell et~al., {\it {Improved measurement of
  the anti-d / anti-u asymmetry in the nucleon sea}},  {\em Phys. Rev.} {\bf
  D64} (2001) 052002, [\href{http://arxiv.org/abs/hep-ex/0103030}{{\tt
  hep-ex/0103030}}].

\bibitem{Aaltonen:2010zza}
{\bf CDF} Collaboration, T.~A. Aaltonen et~al., {\it {Measurement of
  $d\sigma/dy$ of Drell-Yan $e^+e^-$ pairs in the $Z$ Mass Region from
  $p\bar{p}$ Collisions at $\sqrt{s}=1.96$ TeV}},  {\em Phys. Lett.} {\bf B692}
  (2010) 232--239, [\href{http://arxiv.org/abs/0908.3914}{{\tt
  arXiv:0908.3914}}].

\bibitem{Abazov:2007jy}
{\bf D0} Collaboration, V.~M. Abazov et~al., {\it {Measurement of the shape of
  the boson rapidity distribution for $p \bar{p} \to Z/gamma^* \to e^{+} e^{-}$
  + $X$ events produced at $\sqrt{s}$ of 1.96 TeV}},  {\em Phys. Rev.} {\bf
  D76} (2007) 012003, [\href{http://arxiv.org/abs/hep-ex/0702025}{{\tt
  hep-ex/0702025}}].

\bibitem{Aaltonen:2008eq}
{\bf CDF} Collaboration, T.~Aaltonen et~al., {\it {Measurement of the Inclusive
  Jet Cross Section at the Fermilab Tevatron p anti-p Collider Using a
  Cone-Based Jet Algorithm}},  {\em Phys. Rev.} {\bf D78} (2008) 052006,
  [\href{http://arxiv.org/abs/0807.2204}{{\tt arXiv:0807.2204}}]. [Erratum:
  Phys. Rev.D79,119902(2009)].

\bibitem{Abazov:2013rja}
{\bf D0} Collaboration, V.~M. Abazov et~al., {\it {Measurement of the muon
  charge asymmetry in $p\bar{p}$ $\to$ W+X $\to$ $\mu\nu$ + X events at
  $\sqrt{s}$=1.96 TeV}},  {\em Phys. Rev.} {\bf D88} (2013) 091102,
  [\href{http://arxiv.org/abs/1309.2591}{{\tt arXiv:1309.2591}}].

\bibitem{D0:2014kma}
{\bf D0} Collaboration, V.~M. Abazov et~al., {\it {Measurement of the electron
  charge asymmetry in $\boldsymbol{p\bar{p}\rightarrow W+X \rightarrow e\nu
  +X}$ decays in $\boldsymbol{p\bar{p}}$ collisions at
  $\boldsymbol{\sqrt{s}=1.96}$ TeV}},  {\em Phys. Rev.} {\bf D91} (2015), no.~3
  032007, [\href{http://arxiv.org/abs/1412.2862}{{\tt arXiv:1412.2862}}].
  [Erratum: Phys. Rev.D91,no.7,079901(2015)].

\bibitem{Aad:2011dm}
{\bf ATLAS} Collaboration, G.~Aad et~al., {\it {Measurement of the inclusive
  $W^\pm$ and Z/gamma cross sections in the electron and muon decay channels in
  $pp$ collisions at $\sqrt{s}=7$ TeV with the ATLAS detector}},  {\em Phys.
  Rev.} {\bf D85} (2012) 072004, [\href{http://arxiv.org/abs/1109.5141}{{\tt
  arXiv:1109.5141}}].

\bibitem{Aad:2013iua}
{\bf ATLAS} Collaboration, G.~Aad et~al., {\it {Measurement of the high-mass
  Drell--Yan differential cross-section in pp collisions at $\sqrt(s)$=7 TeV
  with the ATLAS detector}},  {\em Phys. Lett.} {\bf B725} (2013) 223--242,
  [\href{http://arxiv.org/abs/1305.4192}{{\tt arXiv:1305.4192}}].

\bibitem{Aad:2011fp}
{\bf ATLAS} Collaboration, G.~Aad et~al., {\it {Measurement of the Transverse
  Momentum Distribution of $W$ Bosons in $pp$ Collisions at $\sqrt{s}=7$ TeV
  with the ATLAS Detector}},  {\em Phys. Rev.} {\bf D85} (2012) 012005,
  [\href{http://arxiv.org/abs/1108.6308}{{\tt arXiv:1108.6308}}].

\bibitem{Aad:2011fc}
{\bf ATLAS} Collaboration, G.~Aad et~al., {\it {Measurement of inclusive jet
  and dijet production in $pp$ collisions at $\sqrt{s}=7$ TeV using the ATLAS
  detector}},  {\em Phys. Rev.} {\bf D86} (2012) 014022,
  [\href{http://arxiv.org/abs/1112.6297}{{\tt arXiv:1112.6297}}].

\bibitem{Aad:2013lpa}
{\bf ATLAS} Collaboration, G.~Aad et~al., {\it {Measurement of the inclusive
  jet cross section in pp collisions at $\sqrt(s)$=2.76 TeV and comparison to
  the inclusive jet cross section at $\sqrt(s)$=7 TeV using the ATLAS
  detector}},  {\em Eur. Phys. J.} {\bf C73} (2013), no.~8 2509,
  [\href{http://arxiv.org/abs/1304.4739}{{\tt arXiv:1304.4739}}].

\bibitem{ATLAS:2012aa}
{\bf ATLAS} Collaboration, G.~Aad et~al., {\it {Measurement of the cross
  section for top-quark pair production in $pp$ collisions at $\sqrt{s}=7$ TeV
  with the ATLAS detector using final states with two high-pt leptons}},  {\em
  JHEP} {\bf 05} (2012) 059, [\href{http://arxiv.org/abs/1202.4892}{{\tt
  arXiv:1202.4892}}].

\bibitem{ATLAS:2011xha}
{\bf ATLAS} Collaboration, {\it {Measurement of the ttbar production
  cross-section in pp collisions at $\sqrt{s}$ = 7 TeV using kinematic
  information of lepton+jets events}}, .

\bibitem{TheATLAScollaboration:2013dja}
{\bf ATLAS} Collaboration, T.~A. collaboration, {\it {Measurement of the
  $t\bar{t}$ production cross-section in $pp$ collisions at $\sqrt{s}=8$ TeV
  using $e\mu$ events with $b$-tagged jets}}, .

\bibitem{Aad:2015auj}
{\bf ATLAS} Collaboration, G.~Aad et~al., {\it {Measurement of the transverse
  momentum and $\phi ^*_{\eta }$ distributions of Drell–Yan lepton pairs in
  proton–proton collisions at $\sqrt{s}=8$ TeV with the ATLAS detector}},
  {\em Eur. Phys. J.} {\bf C76} (2016), no.~5 291,
  [\href{http://arxiv.org/abs/1512.02192}{{\tt arXiv:1512.02192}}].

\bibitem{Aaboud:2016btc}
{\bf ATLAS} Collaboration, M.~Aaboud et~al., {\it {Precision measurement and
  interpretation of inclusive $W^+$ , $W^-$ and $Z/\gamma ^*$ production cross
  sections with the ATLAS detector}},  {\em Eur. Phys. J.} {\bf C77} (2017),
  no.~6 367, [\href{http://arxiv.org/abs/1612.03016}{{\tt arXiv:1612.03016}}].

\bibitem{Aad:2014kva}
{\bf ATLAS} Collaboration, G.~Aad et~al., {\it {Measurement of the $t\bar{t}$
  production cross-section using $e\mu $ events with b-tagged jets in pp
  collisions at $\sqrt{s}$ = 7 and 8 $\,\mathrm{TeV}$ with the ATLAS
  detector}},  {\em Eur. Phys. J.} {\bf C74} (2014), no.~10 3109,
  [\href{http://arxiv.org/abs/1406.5375}{{\tt arXiv:1406.5375}}]. [Addendum:
  Eur. Phys. J.C76,no.11,642(2016)].

\bibitem{Aaboud:2016pbd}
{\bf ATLAS} Collaboration, M.~Aaboud et~al., {\it {Measurement of the
  $t\bar{t}$ production cross-section using $e\mu$ events with b-tagged jets in
  pp collisions at $\sqrt{s}$=13 TeV with the ATLAS detector}},  {\em Phys.
  Lett.} {\bf B761} (2016) 136--157,
  [\href{http://arxiv.org/abs/1606.02699}{{\tt arXiv:1606.02699}}]. [Erratum:
  Phys. Lett.B772,879(2017)].

\bibitem{Aad:2015mbv}
{\bf ATLAS} Collaboration, G.~Aad et~al., {\it {Measurements of top-quark pair
  differential cross-sections in the lepton+jets channel in $pp$ collisions at
  $\sqrt{s}=8$ TeV using the ATLAS detector}},  {\em Eur. Phys. J.} {\bf C76}
  (2016), no.~10 538, [\href{http://arxiv.org/abs/1511.04716}{{\tt
  arXiv:1511.04716}}].

\bibitem{Aad:2014qja}
{\bf ATLAS} Collaboration, G.~Aad et~al., {\it {Measurement of the low-mass
  Drell-Yan differential cross section at $\sqrt{s}$ = 7 TeV using the ATLAS
  detector}},  {\em JHEP} {\bf 06} (2014) 112,
  [\href{http://arxiv.org/abs/1404.1212}{{\tt arXiv:1404.1212}}].

\bibitem{Aad:2014xaa}
{\bf ATLAS} Collaboration, G.~Aad et~al., {\it {Measurement of the $Z/\gamma^*$
  boson transverse momentum distribution in $pp$ collisions at $\sqrt{s}$ = 7
  TeV with the ATLAS detector}},  {\em JHEP} {\bf 09} (2014) 145,
  [\href{http://arxiv.org/abs/1406.3660}{{\tt arXiv:1406.3660}}].

\bibitem{Chatrchyan:2012xt}
{\bf CMS} Collaboration, S.~Chatrchyan et~al., {\it {Measurement of the
  electron charge asymmetry in inclusive $W$ production in $pp$ collisions at
  $\sqrt{s}=7$ TeV}},  {\em Phys. Rev. Lett.} {\bf 109} (2012) 111806,
  [\href{http://arxiv.org/abs/1206.2598}{{\tt arXiv:1206.2598}}].

\bibitem{Chatrchyan:2013mza}
{\bf CMS} Collaboration, S.~Chatrchyan et~al., {\it {Measurement of the muon
  charge asymmetry in inclusive $pp \to W+X$ production at $\sqrt s =$ 7 TeV
  and an improved determination of light parton distribution functions}},  {\em
  Phys. Rev.} {\bf D90} (2014), no.~3 032004,
  [\href{http://arxiv.org/abs/1312.6283}{{\tt arXiv:1312.6283}}].

\bibitem{Chatrchyan:2013tia}
{\bf CMS} Collaboration, S.~Chatrchyan et~al., {\it {Measurement of the
  differential and double-differential Drell-Yan cross sections in
  proton-proton collisions at $\sqrt{s} =$ 7 TeV}},  {\em JHEP} {\bf 12} (2013)
  030, [\href{http://arxiv.org/abs/1310.7291}{{\tt arXiv:1310.7291}}].

\bibitem{Chatrchyan:2013uja}
{\bf CMS} Collaboration, S.~Chatrchyan et~al., {\it {Measurement of associated
  W + charm production in pp collisions at $\sqrt{s}$ = 7 TeV}},  {\em JHEP}
  {\bf 02} (2014) 013, [\href{http://arxiv.org/abs/1310.1138}{{\tt
  arXiv:1310.1138}}].

\bibitem{Chatrchyan:2013faa}
{\bf CMS} Collaboration, S.~Chatrchyan et~al., {\it {Measurement of the $t
  \bar{t}$ production cross section in the dilepton channel in pp collisions at
  $\sqrt{s}$ = 8 TeV}},  {\em JHEP} {\bf 02} (2014) 024,
  [\href{http://arxiv.org/abs/1312.7582}{{\tt arXiv:1312.7582}}]. [Erratum:
  JHEP02,102(2014)].

\bibitem{Chatrchyan:2012bra}
{\bf CMS} Collaboration, S.~Chatrchyan et~al., {\it {Measurement of the
  $t\bar{t}$ production cross section in the dilepton channel in $pp$
  collisions at $\sqrt{s}=7$ TeV}},  {\em JHEP} {\bf 11} (2012) 067,
  [\href{http://arxiv.org/abs/1208.2671}{{\tt arXiv:1208.2671}}].

\bibitem{Chatrchyan:2012ria}
{\bf CMS} Collaboration, S.~Chatrchyan et~al., {\it {Measurement of the
  $t\bar{t}$ production cross section in $pp$ collisions at $\sqrt{s}=7$ TeV
  with lepton + jets final states}},  {\em Phys. Lett.} {\bf B720} (2013)
  83--104, [\href{http://arxiv.org/abs/1212.6682}{{\tt arXiv:1212.6682}}].

\bibitem{Khachatryan:2016pev}
{\bf CMS} Collaboration, V.~Khachatryan et~al., {\it {Measurement of the
  differential cross section and charge asymmetry for inclusive $\mathrm
  {p}\mathrm {p}\rightarrow \mathrm {W}^{\pm }+X$ production at ${\sqrt{s}} =
  8$ TeV}},  {\em Eur. Phys. J.} {\bf C76} (2016), no.~8 469,
  [\href{http://arxiv.org/abs/1603.01803}{{\tt arXiv:1603.01803}}].

\bibitem{Khachatryan:2015luy}
{\bf CMS} Collaboration, V.~Khachatryan et~al., {\it {Measurement of the
  inclusive jet cross section in pp collisions at $\sqrt{s} = 2.76\,\text
  {TeV}$}},  {\em Eur. Phys. J.} {\bf C76} (2016), no.~5 265,
  [\href{http://arxiv.org/abs/1512.06212}{{\tt arXiv:1512.06212}}].

\bibitem{Khachatryan:2016mqs}
{\bf CMS} Collaboration, V.~Khachatryan et~al., {\it {Measurement of the t-tbar
  production cross section in the e-mu channel in proton-proton collisions at
  $sqrt(s)$ = 7 and 8 TeV}},  {\em JHEP} {\bf 08} (2016) 029,
  [\href{http://arxiv.org/abs/1603.02303}{{\tt arXiv:1603.02303}}].

\bibitem{Khachatryan:2015oqa}
{\bf CMS} Collaboration, V.~Khachatryan et~al., {\it {Measurement of the
  differential cross section for top quark pair production in pp collisions at
  $\sqrt{s} = 8\,\text {TeV} $}},  {\em Eur. Phys. J.} {\bf C75} (2015), no.~11
  542, [\href{http://arxiv.org/abs/1505.04480}{{\tt arXiv:1505.04480}}].

\bibitem{Khachatryan:2015oaa}
{\bf CMS} Collaboration, V.~Khachatryan et~al., {\it {Measurement of the Z
  boson differential cross section in transverse momentum and rapidity in
  proton–proton collisions at 8 TeV}},  {\em Phys. Lett.} {\bf B749} (2015)
  187--209, [\href{http://arxiv.org/abs/1504.03511}{{\tt arXiv:1504.03511}}].

\bibitem{Aaij:2012vn}
{\bf LHCb} Collaboration, R.~Aaij et~al., {\it {Inclusive $W$ and $Z$
  production in the forward region at $\sqrt{s} = 7$ TeV}},  {\em JHEP} {\bf
  06} (2012) 058, [\href{http://arxiv.org/abs/1204.1620}{{\tt
  arXiv:1204.1620}}].

\bibitem{Aaij:2012mda}
{\bf LHCb} Collaboration, R.~Aaij et~al., {\it {Measurement of the
  cross-section for $Z \to e^+e^-$ production in $pp$ collisions at
  $\sqrt{s}=7$ TeV}},  {\em JHEP} {\bf 02} (2013) 106,
  [\href{http://arxiv.org/abs/1212.4620}{{\tt arXiv:1212.4620}}].

\bibitem{Chatrchyan:2012bja}
{\bf CMS} Collaboration, S.~Chatrchyan et~al., {\it {Measurements of
  differential jet cross sections in proton-proton collisions at $\sqrt{s}=7$
  TeV with the CMS detector}},  {\em Phys. Rev.} {\bf D87} (2013), no.~11
  112002, [\href{http://arxiv.org/abs/1212.6660}{{\tt arXiv:1212.6660}}].
  [Erratum: Phys. Rev.D87,no.11,119902(2013)].

\bibitem{Aaij:2015gna}
{\bf LHCb} Collaboration, R.~Aaij et~al., {\it {Measurement of the forward $Z$
  boson production cross-section in $pp$ collisions at $\sqrt{s}=7$ TeV}},
  {\em JHEP} {\bf 08} (2015) 039, [\href{http://arxiv.org/abs/1505.07024}{{\tt
  arXiv:1505.07024}}].

\bibitem{Aaij:2015zlq}
{\bf LHCb} Collaboration, R.~Aaij et~al., {\it {Measurement of forward W and Z
  boson production in $pp$ collisions at $\sqrt{s}=8 $ TeV}},  {\em JHEP} {\bf
  01} (2016) 155, [\href{http://arxiv.org/abs/1511.08039}{{\tt
  arXiv:1511.08039}}].

\bibitem{Patrignani:2016xqp}
{\bf Particle Data Group} Collaboration, C.~Patrignani et~al., {\it {Review of
  Particle Physics}},  {\em Chin. Phys.} {\bf C40} (2016), no.~10 100001.

\bibitem{Forte:2010ta}
S.~Forte, E.~Laenen, P.~Nason, and J.~Rojo, {\it {Heavy quarks in
  deep-inelastic scattering}},  {\em Nucl. Phys.} {\bf B834} (2010) 116--162,
  [\href{http://arxiv.org/abs/1001.2312}{{\tt arXiv:1001.2312}}].

\bibitem{Ball:2015tna}
R.~D. Ball, V.~Bertone, M.~Bonvini, S.~Forte, P.~Groth~Merrild, J.~Rojo, and
  L.~Rottoli, {\it {Intrinsic charm in a matched general-mass scheme}},  {\em
  Phys. Lett.} {\bf B754} (2016) 49--58,
  [\href{http://arxiv.org/abs/1510.00009}{{\tt arXiv:1510.00009}}].

\bibitem{deFlorian:2016spz}
{\bf LHC Higgs Cross Section Working Group} Collaboration, D.~de~Florian
  et~al., {\it {Handbook of LHC Higgs Cross Sections: 4. Deciphering the Nature
  of the Higgs Sector}},  \href{http://arxiv.org/abs/1610.07922}{{\tt
  arXiv:1610.07922}}.

\bibitem{Ball:2016neh}
{\bf NNPDF} Collaboration, R.~D. Ball, V.~Bertone, M.~Bonvini, S.~Carrazza,
  S.~Forte, A.~Guffanti, N.~P. Hartland, J.~Rojo, and L.~Rottoli, {\it {A
  Determination of the Charm Content of the Proton}},  {\em Eur. Phys. J.} {\bf
  C76} (2016), no.~11 647, [\href{http://arxiv.org/abs/1605.06515}{{\tt
  arXiv:1605.06515}}].

\bibitem{Ball:2010de}
R.~D. Ball, L.~Del~Debbio, S.~Forte, A.~Guffanti, J.~I. Latorre, J.~Rojo, and
  M.~Ubiali, {\it {A first unbiased global NLO determination of parton
  distributions and their uncertainties}},  {\em Nucl. Phys.} {\bf B838} (2010)
  136--206, [\href{http://arxiv.org/abs/1002.4407}{{\tt arXiv:1002.4407}}].

\bibitem{Baldit:1994jk}
{\bf NA51} Collaboration, A.~Baldit et~al., {\it {Study of the isospin symmetry
  breaking in the light quark sea of the nucleon from the Drell-Yan process}},
  {\em Phys. Lett.} {\bf B332} (1994) 244--250.

\bibitem{Hawker:1998ty}
{\bf NuSea} Collaboration, E.~A. Hawker et~al., {\it {Measurement of the light
  anti-quark flavor asymmetry in the nucleon sea}},  {\em Phys. Rev. Lett.}
  {\bf 80} (1998) 3715--3718, [\href{http://arxiv.org/abs/hep-ex/9803011}{{\tt
  hep-ex/9803011}}].

\bibitem{Garvey:2001yq}
G.~T. Garvey and J.-C. Peng, {\it {Flavor asymmetry of light quarks in the
  nucleon sea}},  {\em Prog. Part. Nucl. Phys.} {\bf 47} (2001) 203--243,
  [\href{http://arxiv.org/abs/nucl-ex/0109010}{{\tt nucl-ex/0109010}}].

\bibitem{Aad:2012sb}
{\bf ATLAS} Collaboration, G.~Aad et~al., {\it {Determination of the strange
  quark density of the proton from ATLAS measurements of the $W \to \ell \nu$
  and $Z \to \ell\ell$ cross sections}},  {\em Phys. Rev. Lett.} {\bf 109}
  (2012) 012001, [\href{http://arxiv.org/abs/1203.4051}{{\tt
  arXiv:1203.4051}}].

\bibitem{Ball:2014uwa}
{\bf NNPDF} Collaboration, R.~D. Ball et~al., {\it {Parton distributions for
  the LHC Run II}},  {\em JHEP} {\bf 04} (2015) 040,
  [\href{http://arxiv.org/abs/1410.8849}{{\tt arXiv:1410.8849}}].

\bibitem{Khalek:2018bbv}
R.~A. Khalek, J.~J. Ethier, and J.~Rojo, {\it {Nuclear Parton Distributions
  from Neural Networks}},  in {\em {Diffraction and Low-x 2018 (Difflowx2018)
  Reggio Calabria, Italy, August 26-September 1, 2018}}, 2018.
\newblock \href{http://arxiv.org/abs/1811.05858}{{\tt arXiv:1811.05858}}.

\bibitem{Pearson:2018tim}
R.~L. Pearson and C.~Voisey, {\it {Towards parton distribution functions with
  theoretical uncertainties}},  in {\em {21st High-Energy Physics International
  Conference in Quantum Chromodynamics (QCD 18) Montpellier, France, July 2-6,
  2018}}, 2018.
\newblock \href{http://arxiv.org/abs/1810.01996}{{\tt arXiv:1810.01996}}.

\bibitem{Buckley:2014ana}
A.~Buckley, J.~Ferrando, S.~Lloyd, K.~Nordström, B.~Page, M.~Rüfenacht,
  M.~Schönherr, and G.~Watt, {\it {LHAPDF6: parton density access in the LHC
  precision era}},  {\em Eur. Phys. J.} {\bf C75} (2015) 132,
  [\href{http://arxiv.org/abs/1412.7420}{{\tt arXiv:1412.7420}}].

\end{thebibliography}\endgroup

\end{document}